\RequirePackage{snapshot}
\documentclass[
  amssymb, amsmath,
  floatfix,
  reprint,%
  twocolumn,%
  superscriptaddress,%
]{revtex4-1}

\usepackage{graphicx}
\usepackage{caption}
\usepackage{hyperref}
\hypersetup{
     colorlinks=true,
     linkcolor=blue
}

\usepackage[T1]{fontenc} 
\usepackage[utf8]{inputenc}
\usepackage{lmodern}

\usepackage{booktabs}
\usepackage{etoolbox}
\usepackage[separate-uncertainty=true]{siunitx}
\usepackage{upgreek}
\usepackage{chemmacros}
\chemsetup{
 formula = mhchem,
 greek = chemgreek,
 modules = thermodynamics,
 modules = redox,
 modules = reactions
}
\usepackage[symbol]{footmisc}
\renewrobustcmd{\bfseries}{\fontseries{b}\selectfont}
\renewrobustcmd{\boldmath}{}

\def\ISO{[A|0|0]}
\def\DIRAC{[A|11B_f|9B_f]^{216}_{\mathcal{D}irac}}
\def\SUP{[A+20B|0|0]}
\def\FnT{[A|11B_f|9B_f]}
\def\FDE{[A|0|20B_f]}

\begin{document}

\title{Investigating solvent effects on the magnetic properties of molybdate ions (\ce{MoO_{4}^{2-}}) with relativistic embedding}
\author{Loïc Halbert}
\affiliation{Univ. Lille, CNRS, UMR 8523 - PhLAM - Physique des Lasers Atomes et Molécules, F-59000 Lille, France}
\author{Ma\l gorzata Olejniczak}
\affiliation{Centre of New Technologies, University of Warsaw, S. Banacha 2c, 02-097 Warsaw, Poland}
\author{Valérie Vallet}
\affiliation{Univ. Lille, CNRS, UMR 8523 - PhLAM - Physique des Lasers Atomes et Molécules, F-59000 Lille, France}
\author{André Severo Pereira Gomes}
\email{andre.gomes@univ-lille.fr}
\affiliation{Univ. Lille, CNRS, UMR 8523 - PhLAM - Physique des Lasers Atomes et Molécules, F-59000 Lille, France}
\date{\today}

\begin{abstract}
We investigate the ability of mechanical and electronic density functional theory (DFT)-based embedding approaches to describe the solvent effects on nuclear magnetic resonance (NMR) shielding constants of the \ce{^{95}Mo} nucleus in the molybdate ion in aqueous solution. From the description obtained from calculations with two- and four-component relativistic Hamiltonians, we find that for such systems spin-orbit coupling effects are clearly important for absolute shielding values, but for relative quantities a scalar relativistic treatment provides a sufficient estimation of the solvent effects. We find that the electronic contributions to the solvent effects are relatively modest yet decisive to provide a more accurate magnetic response of the system, when compared to reference supermolecular calculations. We analyze the errors in the embedding calculations by statistical methods as well as through a real-space representation of NMR shielding densities, which are shown to provide a clear picture of the physical processes at play.

\end{abstract}
\maketitle

\section{Introduction}

Nuclear magnetic resonance (NMR) spectroscopy measures the interaction between the magnetic moments of nuclei and an applied external field, screened by the electrons of the system. NMR is extremely useful for characterizing molecules and materials since it provides detailed information of the local chemical environment around the responding nuclei, and does so in a non-destructive manner. In addition to that, it probes species in their electronic ground state and, in doing so, provides more direct information on the interactions in that state compared to other techniques.

The screening by the electron cloud of the magnetic interactions between a given atomic isotope $K$ and the applied external magnetic field, is represented by the so-called NMR shielding constant, $\sigma^K$, which can be calculated as a second-order derivative of the energy of the molecular system with respect to the magnetic dipole moment of that nucleus ($\vec{m}_K$) and an external magnetic field ($\vec{B}$) at the zero-perturbation limit:
\begin{equation}
\sigma^K_{\alpha\beta} =  \frac{d^2E}{dB_\alpha dm_{K;\beta}} \biggr|_{\vec{B},\{\vec{m}_A\} = 0}.
\label{eq:shield1}
\end{equation}
In most cases, what is measured in experiments is not $\sigma^K$ but rather a signal relative to a chosen reference species, the chemical shift $\delta^K$, though in recent years experimental advances have revived the interest in determining absolute shielding scales~\cite{aucar-arnmr-96-77-2019,nmr-Jackowski-MRC2011-49-600,jaszunski-pnmrs-67-49-2012}.

NMR spectra can be quite difficult to interpret. The difficulties may come from the complexity due to the size of composition of the system (species in solution, disordered materials, etc.), the broadening of signals due to quadrupole interactions (for nuclei with total angular momentum larger than $I =1/2$) or a combination of all of these~\cite{C7CS00682A}. Because of these difficulties, theoretical modeling has become indispensable for the interpretation of experiments~\cite{doi:10.1021/ic301648s,CHARPENTIER20111,doi:10.1021/acs.accounts.6b00404}. Furthermore, for certain systems exhibiting a wide range of chemical shifts, or long acquisition times, theory is crucial to guide experiments by providing the approximate regions of the spectra in which to search for signals.

In this paper we are interested in exploring the use of quantum embedding~\cite{SeveroPereiraGomes2012,jacob-wcms-2014,wesolowski-chemrev-115-5891-2015,Sun:2016hy} to describe the solvent effects on the NMR shielding constants of the \ce{^{95}Mo} on the molybdate dianion in aqueous solution. The molybdate moiety is arguably the simplest experimentally relevant Mo-containing system that can be studied, and it is often used as an experimental reference system in the determination of \ce{^{95}Mo} chemical shifts (in the form of an aqueous solution of sodium molybdate, \ce{Na2MoO4}~\cite{Nguyen2015}. As such, it can be seen as ideal test system to evaluate theoretical approaches that can be applied to the modeling of other, more complicated species. Examples of more complex systems can be found in different classes of molybdenum oxides containing the \ce{MoO3} and \ce{MoO4} moieties. These oxides are found as component of catalysts~\cite{schrock_catalytic_2008,ISI:000170078900026,doi:10.1021/jacs.6b11220}, as fission products~\cite{WALLEZ2014225}, and also make up materials that can be used for applications ranging from nonlinear optical materials~\cite{ZHANG2020109570} to glasses for vitrification of nuclear waste~\cite{Hand:2005:0017-1050:121}. In most cases, the interest lies in understanding how the Mo atom modifies the properties (optical, mechanical etc) of the starting material, and that is where the characterization by NMR becomes interesting.

For materials and complex systems, often treated with band structure methods, the gauge-including projector augmented wave (GIPAW)~\cite{CHARPENTIER20111,doi:10.1021/acs.accounts.6b00404} approach, coupled to density functional theory (DFT) calculations, is the de facto standard for NMR calculations useful to experimentalists. Quantum embedding, on the other hand, represent a class of theoretical methods, often used in combination to  approaches developed for discrete systems (based on DFT but also on more accurate treatments based upon wavefunction theory, WFT). In it a given system is treated quantum mechanically but on the basis of a set of interacting subsystems, for which one can tune the accuracy of the description of each subsystem (by choosing for instance a DFT or WFT treatment), according to a desired compromise between accuracy and computational cost. As such, one can describe the interactions between the species of interest and its surroundings at a fraction of the cost of applying
standard DFT or WFT approaches to the whole system.

Contrary to other embedding 
schemes~\cite{senn_qm/mm_2009,cui-jpcb-104-3721-2000,frank-jctc-8-1480-2012},
 such as those in which only the changes in structure due to the 
environment are considered (mechanical embedding), or where the 
environment is represented classically 
(QM/MM)~\cite{fukal-pccp-21-9924-2019}, quantum embedding is yet to be extensively used for describing magnetic properties in general (and NMR shieldings in particular), in spite of developments of response theory for magnetic perturbations for the frozen density embedding or subsystem DFT 
methods~\cite{jacob-jcp-125-194104-2006,bulo-jpca-112-2640-2008,jacob-jcc-29-1011-2008,gotz-jcp-140-104107-2014,Olejniczak2017}, as well as the generalizations of partition DFT to the spin-polarized case~\cite{Mosquera2012} and the subsystem formulation of time-dependent current density functional theory~\cite{Mosquera2014}.
This is unfortunate since these approaches have the advantage of allowing, or providing a basis for,  
a seamless combination of different electronic structure approaches 
(DFT and WFT), but also the combination of different Hamiltonians, such as 
those taking into account relativistic effects.

Relativistic effects~\cite{Pyykko1988,dyall-faegri-book-2007,reiher_book,Autschbach2012} (scalar relativistic effects and spin-orbit coupling) are now recognized as important (if not essential) to approach experimental accuracy in electronic structure calculations. While this is undisputed in the case of heavy elements ($Z>31$), it is increasingly the case that the effects of relativity are recognized and accounted for even for light systems (first and second-row atoms) when the molecular properties of interest involve a description of the electronic structure near the nucleus. This is the case of NMR properties~\cite{nmr-book-kaupp,Autschbach2014}.

We expect that quantum embedding, if shown to be reliable, can provide a flexible framework and enable more sophisticated simulations, either by complementing DFT-only approaches such as GIPAW (by easily letting one explore the use of different relativistic Hamiltonians) or by allowing the use of more accurate methods, such as those based on the coupled cluster ansatz~\cite{bartlett:reviewcc,Helgaker2012}, thus paving the way for a quantitative leap in the interpretation of NMR spectra. As the hydrated \ce{MoO4^{2-}} molybdate dianion appears to have well-structured hydrations shells~\cite{Nguyen2015}, we are particularly interested in assessing how well quantum embedding can effectively reduce the size of the explicit quantum model, by accounting for hydration effects.

The paper is organized as follows: in section~\ref{sec:theory} we outline the key theoretical aspects underpinning the work. This is followed in section~\ref{computational-details} by a summary of the computational details of our calculations, with a particular emphasis in detailing which structural models are considered and introducing a shorthand notation to refer to these. In section~\ref{results} we discuss the ability of the different embedding approaches to represent the molybdate ion in solution, the influence of the different relativistic Hamiltonians on the NMR shielding constants, and characterize (including in real space) the errors of the different embedding methods. We conclude and offer perspectives for future work in section~\ref{conclusions}.

\section{Theory}
\label{sec:theory}

In what follows, we employ the standard notation for molecular orbital 
indices (with $i, j, \ldots$ corresponding to occupied, $a, b, \ldots$ - to 
virtual and $p, q, \ldots$ - to general molecular orbitals) and the 
summation convention over repeated indices. We use the SI-based atomic 
units (~$\hbar = m_e = e =1/(4\pi\epsilon_0) = 
1$)~\cite{whiffen-pac-50-75-1978}. Employing the second-quantization 
formalism and an exponential parametrization of the unperturbed, closed-shell ground 
state wave function {$\psi(\boldsymbol{\kappa})$ (with the orbital-rotation operator, $\hat{\kappa} = \sum_{pq} 
\kappa_{pq}p^\dagger q$, acting on a trial wave function $\psi_{0}$}
\begin{align}
\psi(\boldsymbol{\kappa}) &= 
e^{\hat{\kappa}}\psi_{0},
\label{eq:exppar}
\end{align}
{and a parametrization of the perturbed wave function in terms of first-order orbital-rotations ($\hat{\kappa}^{(1)} = \sum_{pq} \kappa^{(1)}_{pq}p^\dagger q$)~\cite{Olejniczak2017}}, the derivative in Eq.~\ref{eq:shield1} can be further expressed as
\begin{equation}
\frac{d^2E}{dB_\alpha dm_{K;\beta}} \biggr|_{\varepsilon = 0} = 
\frac{\partial^2 E }{ \partial \kappa^{(1)}_{pq} \partial m_{K;\beta}}  \frac{\partial \kappa^{(1)}_{pq}}{\partial B_\alpha} \biggr|_{\varepsilon = 0}
+
\frac{\partial^2E}{\partial B_\alpha \partial m_{K;\beta}} \biggr|_{\varepsilon = 0},
\label{eq:shield2}
\end{equation}
where all perturbing fields are collected in $\varepsilon$.
Partial derivatives of the orbital rotation parameters ($\{\kappa^{(1)}_{pq}\}$) with respect to the perturbing field are optimized in the linear response equations, which in the static-field regime can be compactly presented as:
\begin{equation}
E_0^{[2]} X_{\varepsilon_1} = - E_{\varepsilon_1}^{[1]},
\label{eq:lr}
\end{equation}
with $E_0^{[2]}$, $E_{\varepsilon_1}^{[1]}$ and $X_{\varepsilon_1}$ corresponding to the electronic Hessian, the property gradient and the solution vector ($X_{\varepsilon_1} = ( \partial \kappa^{(1)}_{pq} / \partial {\varepsilon_1})_{\varepsilon = 0}$), respectively~\cite{saue-jcp-118-522-2003}.
The two former quantities require the formulation of various derivatives of Fock or Kohn-Sham matrices, therefore they are strictly connected to the formalism employed in the calculations.  

In this work, we exploit the computational model based on relativistic Hamiltonians
~\cite{saue-cpc-12-3077-2011} and on the spin-density functional theory (SDFT)~\cite{barth-jpc-5-1629-1972,rajagopal-prb-7-1912-1973,jacob-ijqc-112-3661-2012}. 
The closed-shell system is described by the electron density ($\rho_0$) and the spin-density calculated in a non-collinear fashion as a norm of the spin-magnetization vector ($s = |\rho_\mu|$, $\mu \in \{x,y,z\}$). Following previous works~\cite{bast-ijqc-109-2091-2009,komorovsky-jcp-128-104101-2008,olejniczak-jcp-136-014108-2012,Olejniczak2017}, we collect the electron density and the spin-magnetization vector into one variable - a general density component, $\rho_k$ ($k \in \{0,x,y,z\}$).

This framework becomes slightly modified in the embedding situation. The DFT-subsystem-based approaches rely on the partitioning of the total system into (interacting) subsystems~\cite{SeveroPereiraGomes2012,goez-in-frontiers-of-quantum-chem-2018,jacob-wcms-2014},
which is realized by expressing the electron density of the full system as a sum of electron densities of subsystems~\cite{wesolowski-jpc-97-8050-1993,wesolowski-cpl-248-71-1996,wesolowski-radft-371-2002,wesolowski-chemrev-115-5891-2015} (the same applies to the spin density if it is included in the formalism),
\begin{equation}
\rho_k^\text{tot} (\vec{r}) = \rho_k^I (\vec{r}) +\rho_k^{II} (\vec{r}) 
\label{eq:FDE_sum_density}
\end{equation} 
Consequently, the total energy may be decomposed into contributions corresponding to energies of subsystems and the interaction energy term which depends on the (spin) densities of all subsystems~\cite{Olejniczak2017},
\begin{equation}
E_\text{tot}[\rho_k^\text{tot}] = E_I[\rho_k^I] + E_{II}[\rho_k^{II}] + E_\text{int}[\rho^{I}_k, \rho^{II}_k].
\label{eq:FDE_sum_energy}
\end{equation} 
The analytical formula for the interaction energy term can be derived from the (S)DFT expression for the energy of the full system and reads as:
\begin{align}
    E_\text{int}[\rho^{I}_k, \rho^{II}_k]
      &=  \int \left[ \rho_0^{I}(\vec{r}) v_\text{nuc}^{II}(\vec{r}) + \rho_0^{II}(\vec{r}) v_\text{nuc}^{I}(\vec{r}) \right]d\vec{r}  \nonumber\\
       & + \iint \frac{\rho_0^{I}(\vec{r}_1) \rho_0^{II}(\vec{r}_2)}{|\vec{r}_1 - \vec{r}_2|} d\vec{r}_1d\vec{r}_2
      + E_\text{nuc}^{I,II}       \nonumber\\
      &+ E_\text{xc}^\text{nadd} + T_{s}^\text{nadd},
      \label{eq:Eint}
\end{align}
where, in addition to terms describing the interaction of the electron density of a subsystem with the electron density and the nuclear potential of another subsystem and a term related to the nuclear repulsion energy between subsystems (first four terms in Eq.~\ref{eq:Eint}), the non-additive exchange-correlation ($E_\text{xc}^\text{nadd}$) and kinetic energy ($T_{s}^\text{nadd}$) contributions emerge. These non-additive contributions~\cite{hofener-jcp-136-044104-2012} are defined as:
  \begin{equation}
    X^\text{nadd} \equiv X^\text{nadd}[\rho_k^{I}, \rho_k^{II}] = X[\rho_k^\text{tot}] - X[\rho_k^{I}] - X[\rho_k^{II}]
    \label{eq:nonadd}
  \end{equation}
and they depend on the general density components of all subsystems.  
In typical FDE calculations, one subsystem is chosen as \emph{active}, while another subsystem constitute its \emph{environment}. In this setting, the \emph{active} density (e.g. $\rho_k^I$) is optimized by solving the so-called Kohn-Sham (KS) equations for a constrained electron density (KSCED)~\cite{Wesolowski1993}, in which an effective KS potential known from a standard (S)DFT is augmented by the embedding potential responsible for describing the interaction of an \emph{active} subsystem with the \emph{environment} (here: $\rho_k^{II}$): 
\begin{align}
    v_\text{emb;k}^I(\vec{r}) &= \frac{\delta E_\text{int}}{\delta \rho_k^I(\vec{r})} 
                   = \frac{\delta E_\text{xc}^\text{nadd}}{\delta \rho_k^I} +  \frac{\delta T_{s}^\text{nadd}}{\delta \rho_k^I}  + v_\text{nuc}^{II}(\vec{r}) \nonumber\\
                   &+ \int \frac{\rho_0^{II}(\vec{r}')}{|\vec{r} - \vec{r}'|}d\vec{r}'
    \label{eq:embpot}.
\end{align}
It is also possible to relax the density of the \emph{environment} by solving the analogous KSCED equations set up for $\rho_k^{II}$. Repeating this procedure in an iterative manner, known as the \emph{freeze-thaw}~\cite{jacob-jcc-29-1011-2008} (FnT) cycle allows to optimize the densities of all subsystems. {Clearly,  the embedding potential in Eq.~\ref{eq:embpot} couples the systems in the ground state.}

At this point it should be noted that although FDE is formally exact, its practical realization requires approximation to the non-additive exchange-correlation and kinetic energies. This is typically done with approximate density functionals, which - especially for the kinetic energy part - have limited accuracy~\cite{SeveroPereiraGomes2012,jacob-wcms-4-325-2014,schluns-pccp-17-14323-2015,fux-jcp-132-164101-2010,artiukhin-jcp-142-234101-2015}.
From the presence of the interaction energy term in the total energy expression (Eq.~\ref{eq:FDE_sum_density}) and from its functional dependence on the densities of all subsystems, it becomes evident that the quantities which in the linear response theory are formulated as various energy derivatives ($E_0^{[2]}$, $E_{\varepsilon_1}^{[1]}$ in Eq.~\ref{eq:lr}) now have to be augmented with analogous derivatives of $E_{int}$. 

These additional contributions, derived by using the chain rule, involve the first- and second-order derivatives of $E_{int}$ with respect to the densities of subsystems - the embedding potential (Eq.~\ref{eq:embpot}) and the embedding kernel (see~\cite{Olejniczak2017} for details). In effect, the property gradient and the electronic Hessian (Eq.~\ref{eq:lr}) are elegantly subdivided into contributions from isolated subsystems and terms related to the interaction input~\cite{hofener-jcp-136-044104-2012}. In practice, these derivatives can be coded analogously to the exchange-correlation terms in standard (S)DFT technique.

There is, however, an additional difficulty in the molecular property calculations, specific to the embedding setting. Namely, one needs to account for the fact that the property of interest arises as a response of the full system - comprising of all subsystems - to certain external perturbations.In particular, for the NMR shielding tensor of a nucleus K this can be symbolically expressed as
\begin{equation}
\sigma_{\alpha\beta}^K = \sigma_{\alpha\beta}^{K, I} + \sigma_{\alpha\beta}^{K, II}.
\label{eq:shield_fde}
\end{equation}
{As evoked above and discussed in~\cite{Olejniczak2017}, the response of the interaction term will introduce terms in the electronic Hessian and property gradient that couple the response of the subsystems to the external perturbations, so each $\sigma_{\alpha\beta}^{K, I}$ and $\sigma_{\alpha\beta}^{K, II}$ formally contain intra-subsystem and inter-subsystem contributions. In this work we consider the same approximation as in prior works~\cite{jacob-jcp-125-194104-2006,bulo-jpca-112-2640-2008} and disregard the inter-subsystem (coupling) contributions in the calculation of $\sigma_{\alpha\beta}^{K, I}$ and $\sigma_{\alpha\beta}^{K, II}$.}

{We note that the term $\sigma_{\alpha\beta}^{K, II}$, assuming that the nucleus K belongs to subsystem I}, can be interpreted as an effect that the perturbation due to $B_\alpha$ in subsystem II has in the point of the center of nucleus K - an idea analogous to the concept of the nucleus independent chemical shift (NICS)~\cite{schleyer-jacs-118-6317-1996}.
A final complication in the formulation of the NMR shielding tensor arises as a consequence of the dependence of this property operator (more specifically - the Zeeman term) on the choice of the gauge origin, which leads to unphysical results in incomplete basis set regime. The typical remedy is to use the London atomic orbitals (LAOs)~\cite{london-jpr-8-397-1937,helgaker-chemrev-99-293-1999} instead of atomic orbitals, however with the price of the need to calculate additional terms in the property gradient~\cite{ilias-jcp-131-124119-2009,olejniczak-jcp-136-014108-2012,Olejniczak2017}.

The NMR shielding tensor (Eqs.~\ref{eq:shield1},~\ref{eq:shield2}) can also be determined from numerical integration of the NMR shielding density~\cite{jameson-jcp-73-5684-1980,jameson-jpc-83-3366-1979}. This property density can be calculated (analytically) in real space by using the relation to the magnetically-induced current density, $j^B$ - also a second-order tensor quantity, which in the relativistic framework~\cite{bast-cp-356-187-2009,sulzer-pccp-13-20682-2011} reads
\begin{equation}
j^B = \left. \frac{d\vec{j}}{d\vec{B}} \right|_{\vec{B} = 0}; 
\quad
\vec{j} = -e \psi_i^\dagger  c \vec{\alpha} \psi_i.
\label{eq:j^B_j}
\end{equation}
The NMR shielding density is an integrand in the following expression:
\begin{equation}
\sigma^K_{\alpha\beta} = -\frac{1}{c^2} \int \frac{1}{r_K^3} \left( \vec{r}_K \times \vec{j}^{\,B_\alpha} \right)_\beta d\vec{r}.
\label{eq:shield3}
\end{equation}
This reformulation of $\sigma^K_{\alpha\beta}$ has multiple advantages. First, it allows to recalculate the NMR shielding values (therefore serves as a test of a fully analytical approach), secondly it opens the possibility to visualize the NMR shielding density on meshes, which has already proved to be an invaluable asset in the post-production analysis process~\cite{Olejniczak2017}. 
Additional advantage of this formulation is that the NMR shielding density of a nucleus K can easily be calculated in any point in space, even if that point and the center of nucleus K belong to different subsystems, what permits to evaluate the second term of Eq.~\ref{eq:shield_fde}~\cite{jacob-jcp-125-194104-2006}.

\section{Computational details\label{computational-details}}

The solvated molybdate structures used in this work are taken from the Car-Parrinello molecular dynamics (CPMD~\cite{Hutter:2011gl}) calculations of Nguyen and coworkers~\cite{Nguyen2015}. We considered 517 snapshots, each containing the molybdate ion and 20 water molecules. For additional details, the reader is referred to the aforementioned paper. From these we defined four basic structural models: 
\begin{itemize}
\item a supermolecular system, containing the molybdate ion and all water molecules;

\item a subsystem embedding model, where the molybdate ion (and depending on the case, selected water molecules) is the active subsystem and the (remaining) water molecules make up the environment, but in which a given number of water molecules nearest to the active subsystem are relaxed through \emph{freeze-thaw} cycles while the others remain frozen. We will refer to this model as FnT. 

The electron densities and electrostatic potential for the environment are obtained either from calculations on individual water molecules (and we refer to a ``fragmented'' environment) or by these forming a single subsystem (and we refer to a ``grouped'' environment);

\item a frozen density embedding model, in which all water molecules in the environment are kept frozen. We will refer to this model as FDE, and the same notations for the environment (``fragmented'' or ``grouped'') as for FnT is used;

\item an isolated molybdate ion. As the structures of the latter are not obtained in the gas phase but rather through the CPMD calculations, this model is equivalent to that of mechanical embedding, and as such we will refer to it as ``mechanical embedding''.

\end{itemize}

Using the labels ``A'' for the molybdate ion and ``B'' for the water molecules, and indices \textbf{g} and \textbf{f} to denote the ``\textbf{g}rouped'' or ``\textbf{f}ragmented'' environment, we shall use the following shorthand notation to represent these four cases: $\left[Active\,|\, Relaxed  \, |\, Frozen \right]$. This means, for example, that the supermolecular model is denoted by $\left[A + 20 B | 0 | 0 \right]$ (all species are in the active fragment), the mechanical embedding by $\left[A  | 0 | 0 \right]$, and a FnT model in which 20 molecules make up the environment but in which 11 are relaxed (and all are taken as individual fragments), is denoted by $\left[A | 11 B_f | 9 B_f \right]$.

The execution of all calculations, including the preparation of the different fragments, has been handled with the PyADF~\cite{Jacob2011} scripting framework.
In all calculations, we employ Becke integration grids, and a gaussian nuclear charge distribution~\cite{visscher-adandt-67-207-1997} to describe the finite nuclear volume~\cite{Andrae2000}, PBE~\cite{perdew-prl-77-3865-1996,perdew-prl-78-1396-1997} exchange-correlation functional. For subsystem embedding and FDE embedding calculations (see below), we have employed the PW91k~\cite{lembarki-pra-50-5328-1994} and PBE~\cite{perdew-prl-77-3865-1996,perdew-prl-78-1396-1997} functionals for calculating the non-additive kinetic and exchange-correlation contributions, respectively, and used a monomer expansion for the basis sets. In the case of subsystem embedding, we have performed 5 \emph{freeze-thaw} cycles. In the NMR calculations we employ London orbitals.

\subsection{One- and two-component calculations}

The scalar relativistic (SR) and spin-orbit (SO) ZORA~\cite{vanlenthe-jcp-105-6505-1996,vanlenthe-jcp-101-9783-1994,vanlenthe-ijqc-57-281-1996,vanlenthe-jcp-110-8943-1999} calculations have been performed with the ADF~\cite{SCMADF} code, using the TZ2P basis sets~\cite{vanlenthe-jcc-24-1142-2003}. The NMR calculations have been performed with the NMR program. 

\subsection{Four-component calculations}

The calculations employing the Dirac-Coulomb (DC) Hamiltonian have been performed with the Dirac code~\cite{DIRAC18}, revision \texttt{54ab939}. For the Mo atom the dyall.cv3z basis~\cite{basis-Dyall-TCA2007-117-483} is used, and for all lighter atoms the aug-cc-pVTZ~\cite{basis-Dunning-JCP1989-90-1007} basis is used. To aid in the convergence of the SCF procedure, in all calculations we have used the atomic start procedure, described in the DIRAC documentation.

For the subsystem embedding and FDE embedding calculations NMR calculations, we remain in the uncoupled response approximation, but take into account contributions from the spin density according to Ref.~\citenum{Olejniczak2017}. We note that unless otherwise noted, the embedding calculations make use of results of embedded ZORA calculations with ADF (embedding potentials, electrostatic potentials and frozen electron densities and gradients of the density for the environment).

The NMR shielding densities used in the analysis of the embedding methods (see section~\ref{analysis-shielding}) have been obtained with the analysis module of the Dirac code, and are based on four-component calculations with the DC Hamiltonian and equation~\ref{eq:shield3}. In the case of embedded calculations, following equation~\ref{eq:shield_fde}, it was necessary to also perform calculations for the environment using the position of the Mo atom as the $\vec{r}_K$ position.

\subsection{Plotting}

All graphs have been prepared with Matplotlib~\cite{Hunter_2007} python library, with the exception of the volumetric density and NMR shielding density plots, for which the  Mayavi~\cite{ramachandran2011mayavi} python library has been used.

\section{Results and discussion\label{results}}

In what follows we are interested in the isotropic part of the NMR shielding tensor of the \ce{^{95}Mo} nucleus, $\sigma_{iso}$, which is 
defined as:
\begin{eqnarray}
\sigma_{iso}=1/3(\sigma_{11}+\sigma_{22}+\sigma_ {33})
\end{eqnarray}
where the indices refer to the principal axes of the tensor according to the Mason's notation~\cite{pmid7804782}, with $\sigma_{11}\leq \sigma_{22} \leq \sigma_{33} $. 

\subsection{Minimal structural model}

An embedding-based computational strategy takes its most effective form, from a computational cost perspective, when the smallest possible molecular moiety can be considered as the active subsystem, and the rest of the system as the environment. In this work such a situation would correspond, as discussed above, to having only the \ce{MoO4^{2-}} ion in the active subsystem.  In this subsection we explore this model. {As our goal in this paper is to investigate the performance of embedding models, we use as reference for comparison the supermolecular model $\left[A + 20 B | 0 | 0 \right]$, since that lets us directly trace back any discrepancies to the approximations introduced by the embedded models.}

\subsubsection{Structural and electronic effects on NMR shieldings}

\begin{figure}
\centering
       	\includegraphics[width=\linewidth]{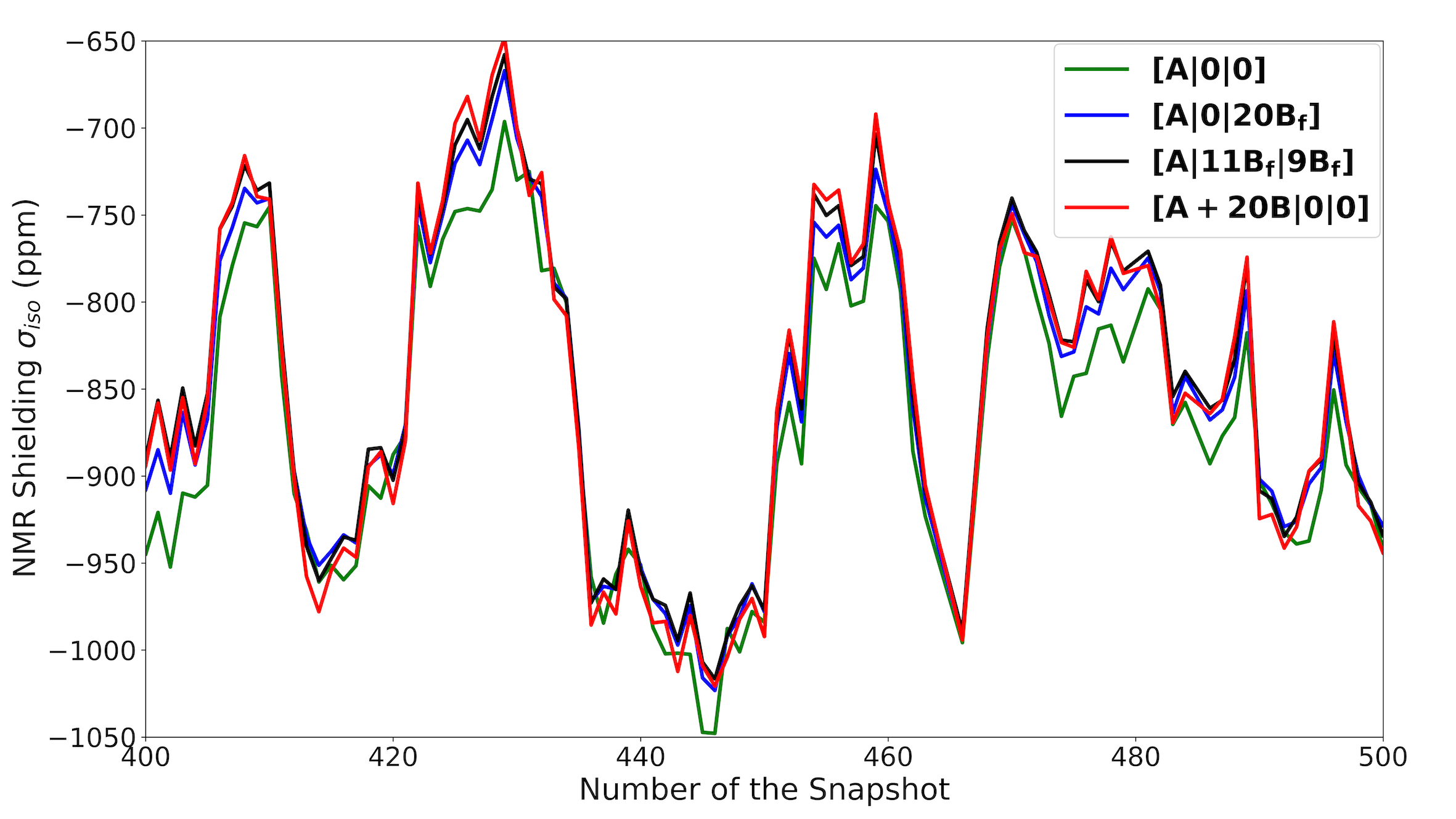}
	\caption{$\sigma_{iso}$ (in ppm) for a subset of the snapshots considered (snapshots 400 to 500), obtained from SR-ZORA calculations for the	
	   supermolecule ($\SUP$), electronic (FnT:$\FnT$ and FDE:$\FDE$) and mechanical ($\ISO$) embedding.}
	\label{fig:Evolution_Shielding_snap400_500}
\end{figure}

We start the discussion with the analysis of the influence of the geometry of molecular complexes on $\sigma_{iso}$ based on the SR-ZORA calculations. This is summarized in Figure~\ref{fig:Evolution_Shielding_snap400_500}, which shows the values of $\sigma_{iso}$ obtained for a subset of 100 geometries out of the 517 considered MD simulations.

An interesting observation that emerges from this figure is that the description offered by different approximations (FnT, FDE, mechanical embedding) is not consistent for all the snapshots. While for certain geometries the electronic part of embedding is an essential component of the model in order to approach the supermolecular $\sigma_{iso}$ values, for others these contributions are negligible and the mechanical embedding is sufficient to reproduce the supermolecular baseline. 

\begin{table}[h]
\centering
\begin{tabular}{
        llc
}
\hline
{Model} & {$\bar{\sigma}_{iso} \pm std$} & {$\Delta\bar{\sigma}_{iso}$ wrt. $\SUP$}\\
\hline
{$\SUP$} & $-848.5\pm108.9$ & {-}\\
{$\FnT$} & $-841.5\pm103.9$& $7.0\pm8.5$\\
{$\FDE$} & $-848.0\pm101.8$& $3.5\pm13.6$\\
{$\ISO$} & $-856.5\pm99.1$ & $-7.9\pm24.5$\\
\hline
\end{tabular}
\caption{Mean ($\bar{\sigma}_{iso}$) and Standard Deviations (\emph{std}) of $\sigma_{iso}$ (in ppm) obtained from all snapshots for SR-ZORA calculations for various models: Supermolecule $\SUP$, FnT $\FnT$, FDE $\FDE$ and isolated $\ISO$. $\Delta\bar{\sigma}_{iso}$ are the deviations with respect to $\SUP$.}
\label{tab:table_all_results_ZORASR}
\end{table}

{The fact that the values of $\sigma_{iso}$ are often close to each other for the different snapshots as shown in Figure~\ref{fig:Evolution_Shielding_snap400_500} is reflected on the similarity of the  mean values of the \ce{^{95}Mo} NMR shielding constants, featured as $\bar{\sigma}_{iso}$ in Table~\ref{tab:table_all_results_ZORASR}}, calculated over all the snapshots for all the considered approximations (FnT, FDE, mechanical embedding).  However, the choice of the computational model does affect the standard deviation of the $\sigma_{iso}$ values distribution - in this case the variance of the NMR shielding values over all the snapshots is significantly smaller if electronic embedding is included in the calculations. This signifies that the data from electronic embedding calculation is statistically less disperse, therefore in this case these can be considered as more accurate for the modeling of  $\sigma_{iso}$ {and $\bar{\sigma}_{iso}$}.

\begin{figure}[htp]
        \centering
	\begin{minipage}{0.90\linewidth}\centering
	   \includegraphics[width=1\linewidth]{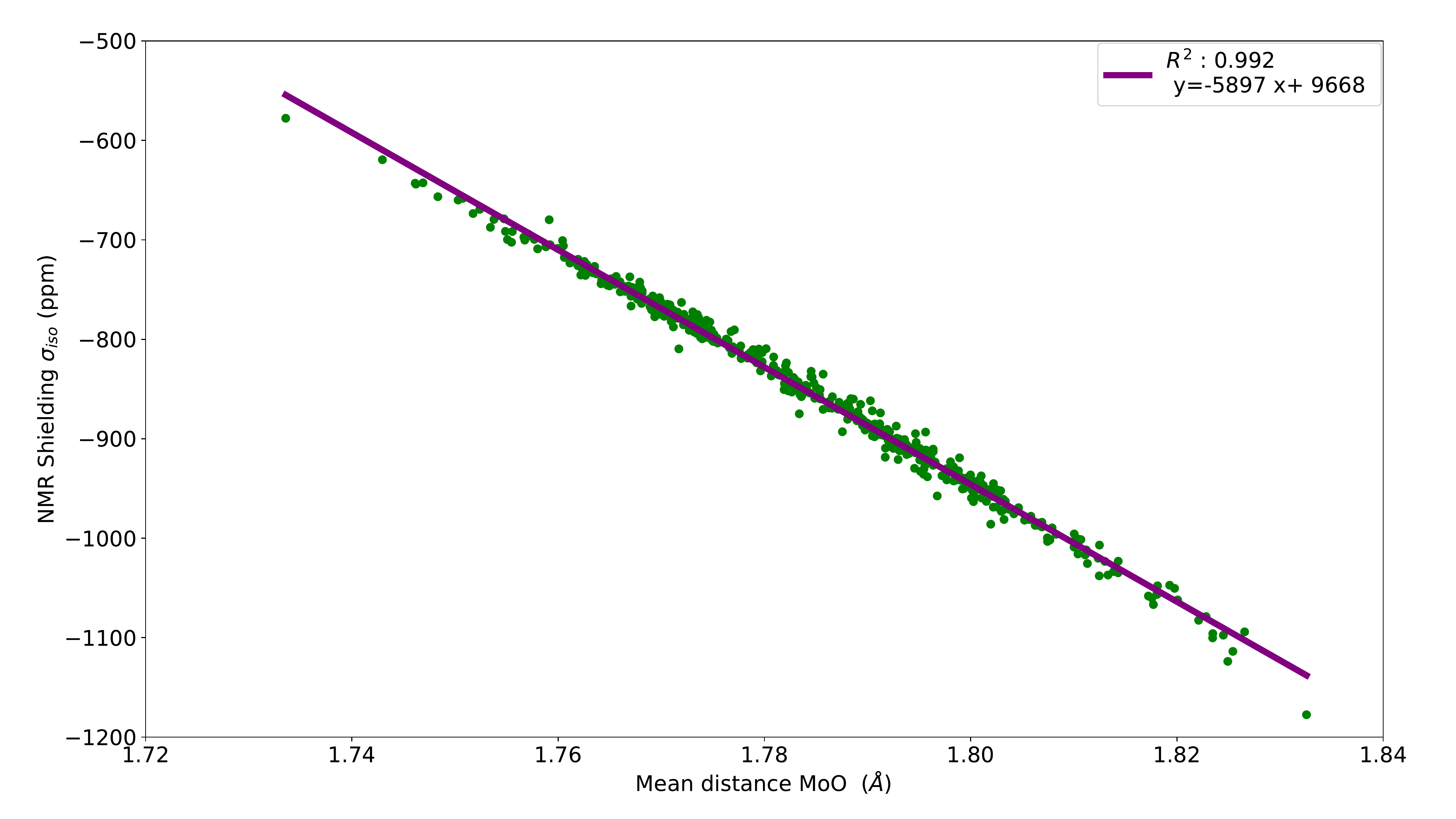}
	     \caption*{{(a)}}
	   \end{minipage}
	\begin{minipage}{0.90\linewidth}\centering
	   \includegraphics[width=1\linewidth]{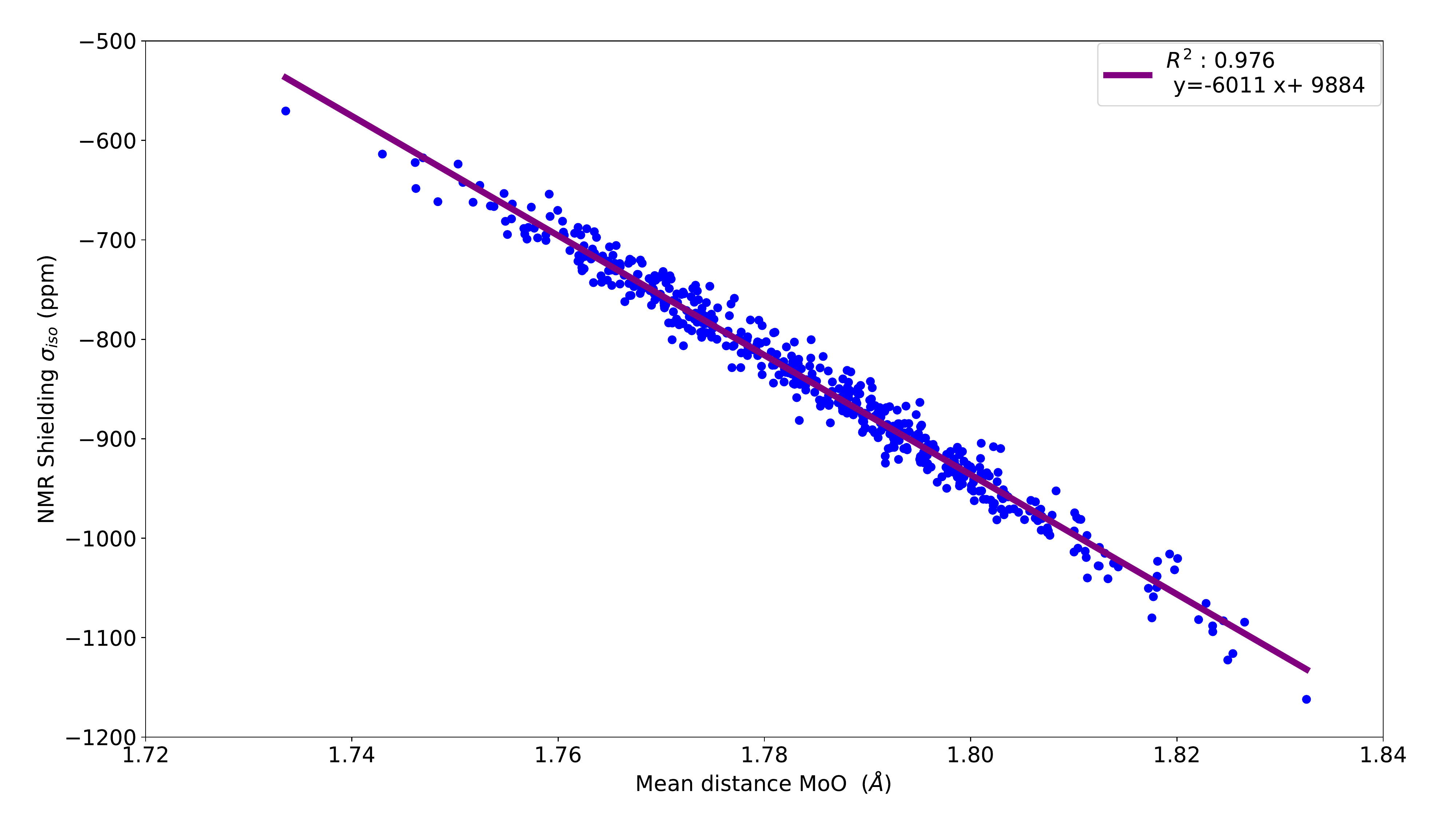}
	   \caption*{{(b)}}
	   \end{minipage}
	\begin{minipage}{0.90\linewidth}\centering
	   \includegraphics[width=1\linewidth]{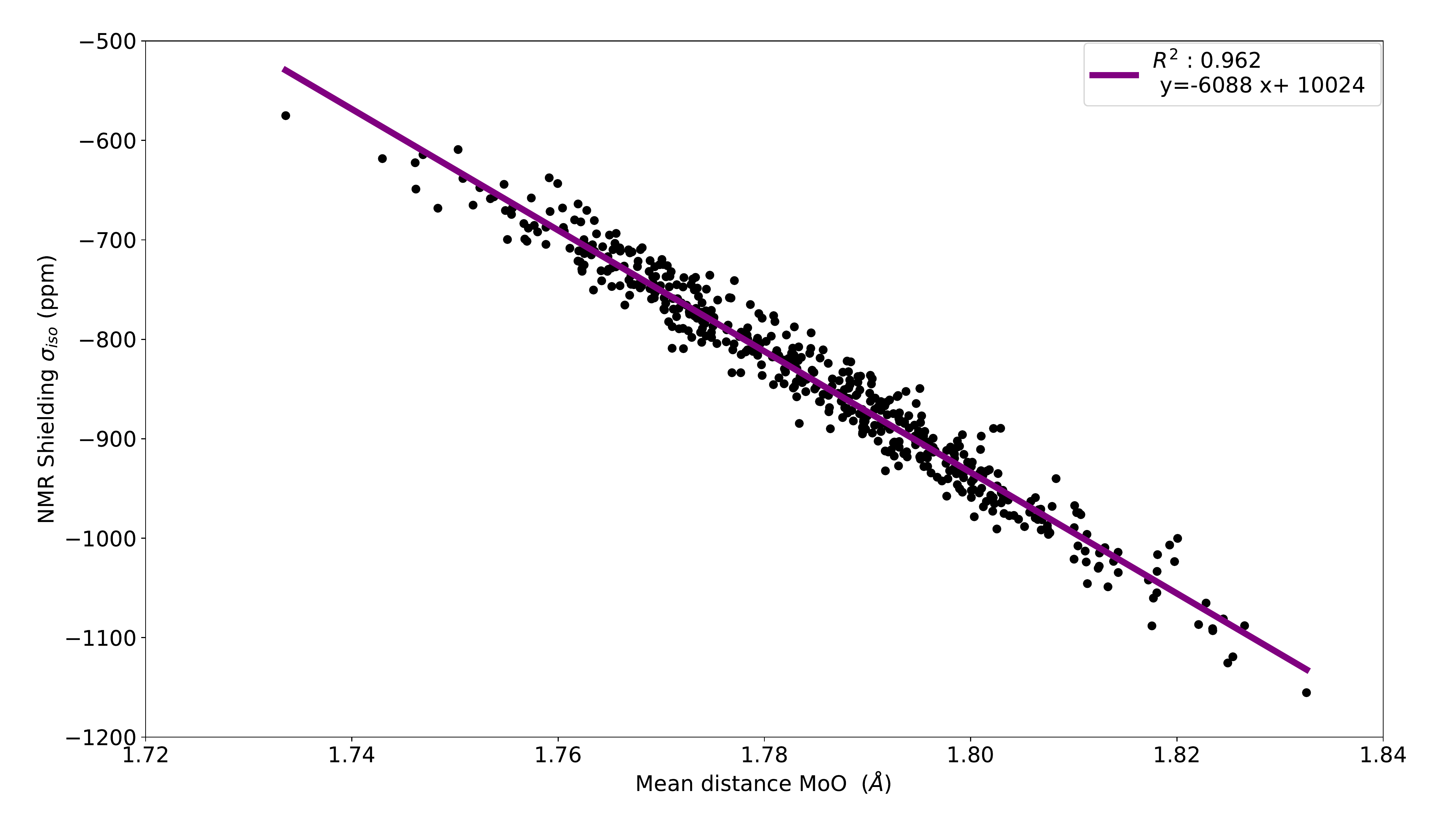}
	   \caption*{{(c)}}
	   \end{minipage}
	\begin{minipage}{0.90\linewidth}\centering
	   \includegraphics[width=1\linewidth]{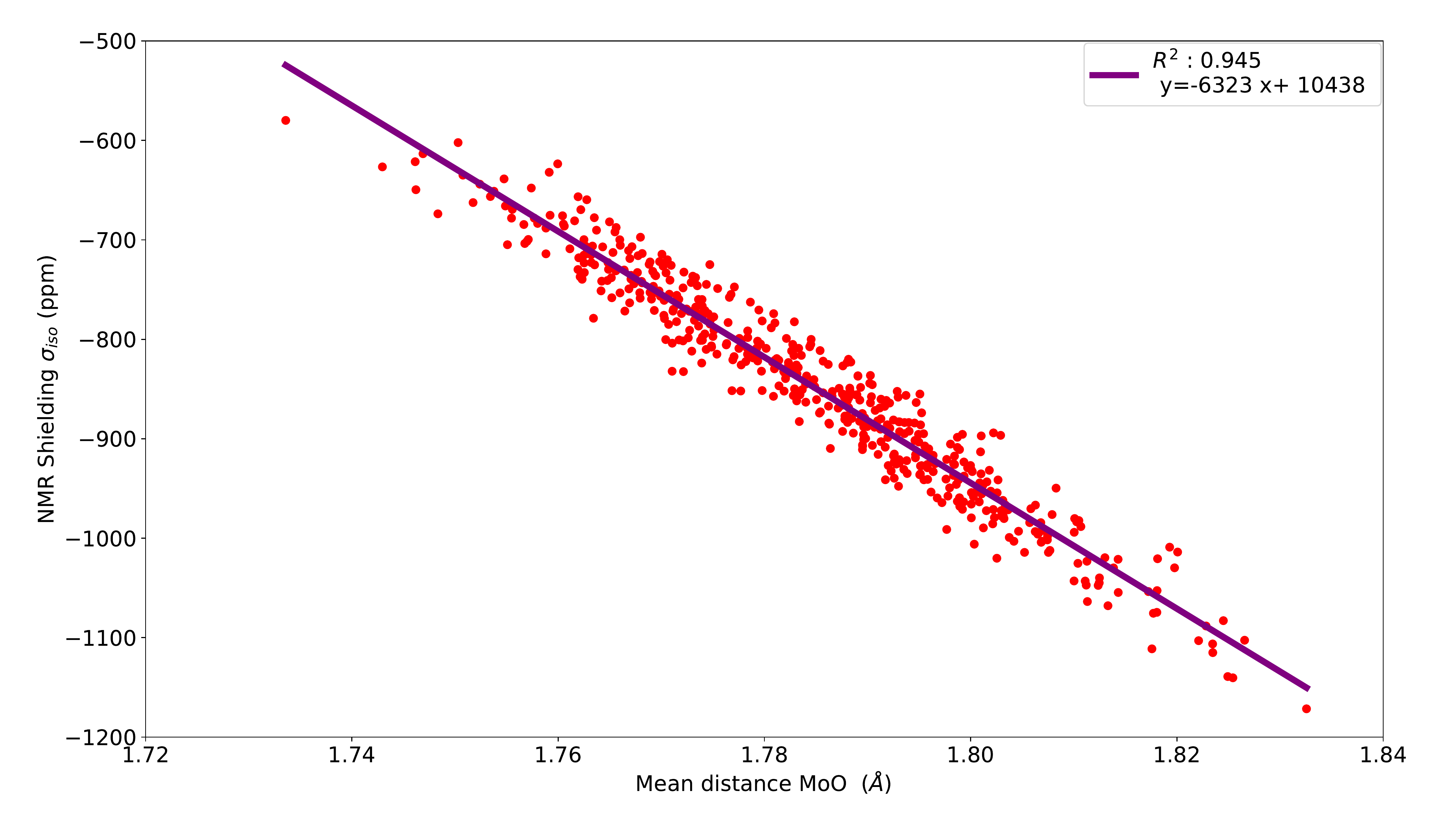}
	   \caption*{{(d)}}
	   \end{minipage}
	\caption{Correlations between $\sigma_{iso}$ (in ppm) and the mean distance between Mo and the oxygen atoms in the Molybdate ion ($\bar{r}$[Mo-O], in \r{A}), obtained from SR-ZORA calculations for various models: (a) isolated $\ISO$; (b) FDE $\FDE$; (c) FnT $\FnT$; and (d) Supermolecule $\SUP$ for all snapshots.}
	\label{fig:ZORASR_vs_meandist}
\end{figure}

In order to understand these structural dependencies of $\sigma_{iso}$ values, we decided to define a measure, the mean distance between the centers of Mo atom and O atoms in \ce{MoO4^{2-}}, hereafter referred to as $\bar{r}$[Mo-O], and to plot the $\sigma_{iso}$ vs $\bar{r}$[Mo-O] in Figure~\ref{fig:ZORASR_vs_meandist}.

The first conclusion of this figure is that there is a significant correlation between the $\sigma_{iso}$  and  $\bar{r}$[Mo-O] for all approximations, as seen from the large values of $R^2$.
The largest correlation between these two quantities was found for the mechanical coupling ($R^2 = 0.992$) and the lowest for the supermolecular results ($R^2 = 0.945$), which indicates the possibility that the effect of the interaction between two subsystem on $\sigma_{iso}$ values cannot be simply attributed to the structural variations, or at least the ones represented by the $\bar{r}$[Mo-O] measure. The FDE and FnT methods recover some of the $\bar{r}$[Mo-O]-dependent contributions, as illustrated by the correlation coefficients ($R^2 = 0.962$ and $0.976$ for FnT and FDE, respectively). The slope of the linear regression also shows a better agreement between supermolecule and electronic embedding calculations than between supermolecule and mechanical embedding.

However, the linearity of the relationship between   $\sigma_{iso}$  and  $\bar{r}$[Mo-O] is interesting since the NMR shielding tensor is considered to be a local property. Yet, as we shall discuss below, there might be other long-range effects that dampen the $r^{-2}$ dependence of the corresponding property operator.

\begin{figure}
        \centering
	\begin{minipage}{0.90\linewidth}\centering
	   \includegraphics[width=1\linewidth]{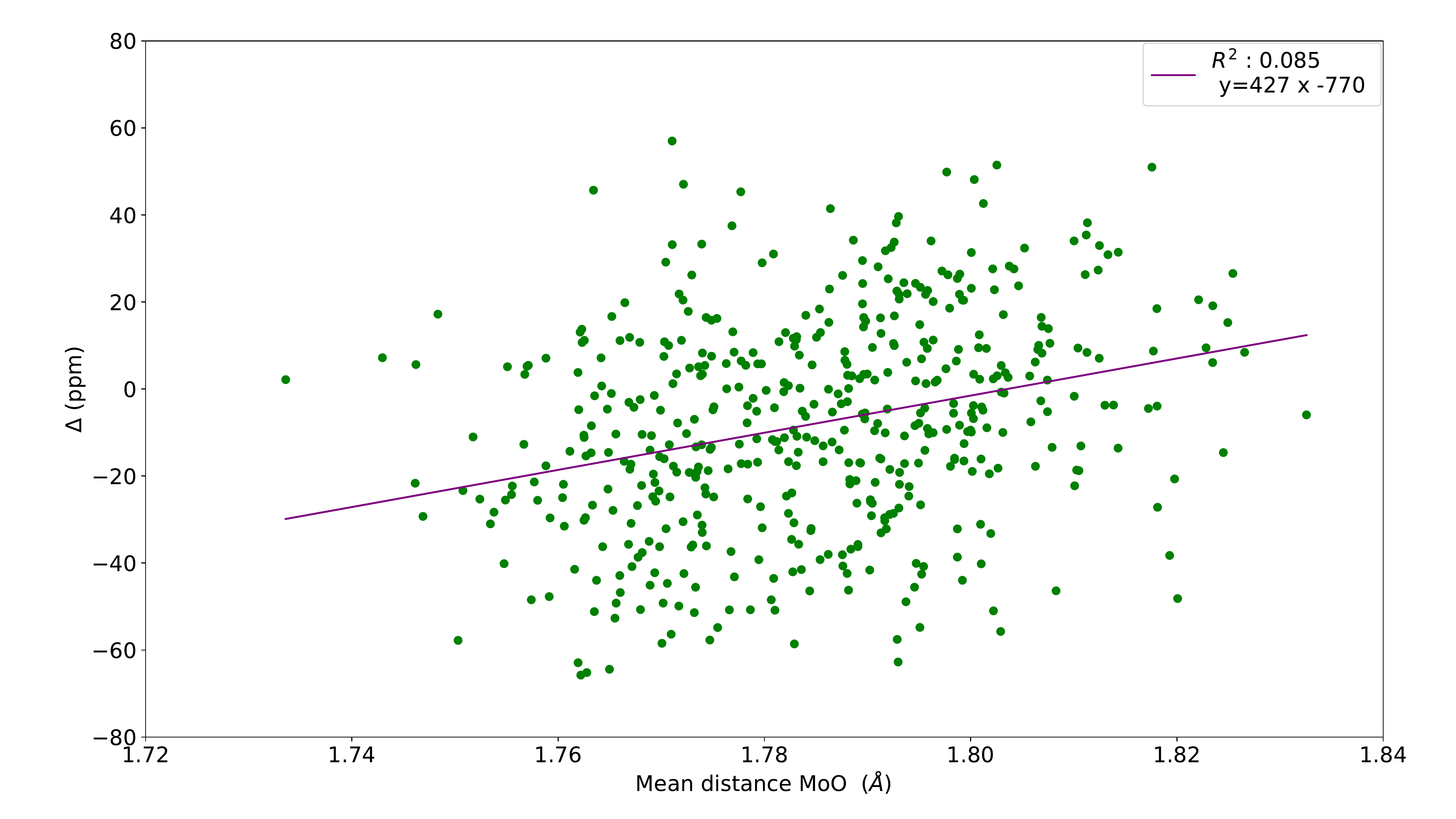}
	   	   
	    \caption*{(a)}
	   \end{minipage}
	\begin{minipage}{0.90\linewidth}\centering
	   \includegraphics[width=1\linewidth]{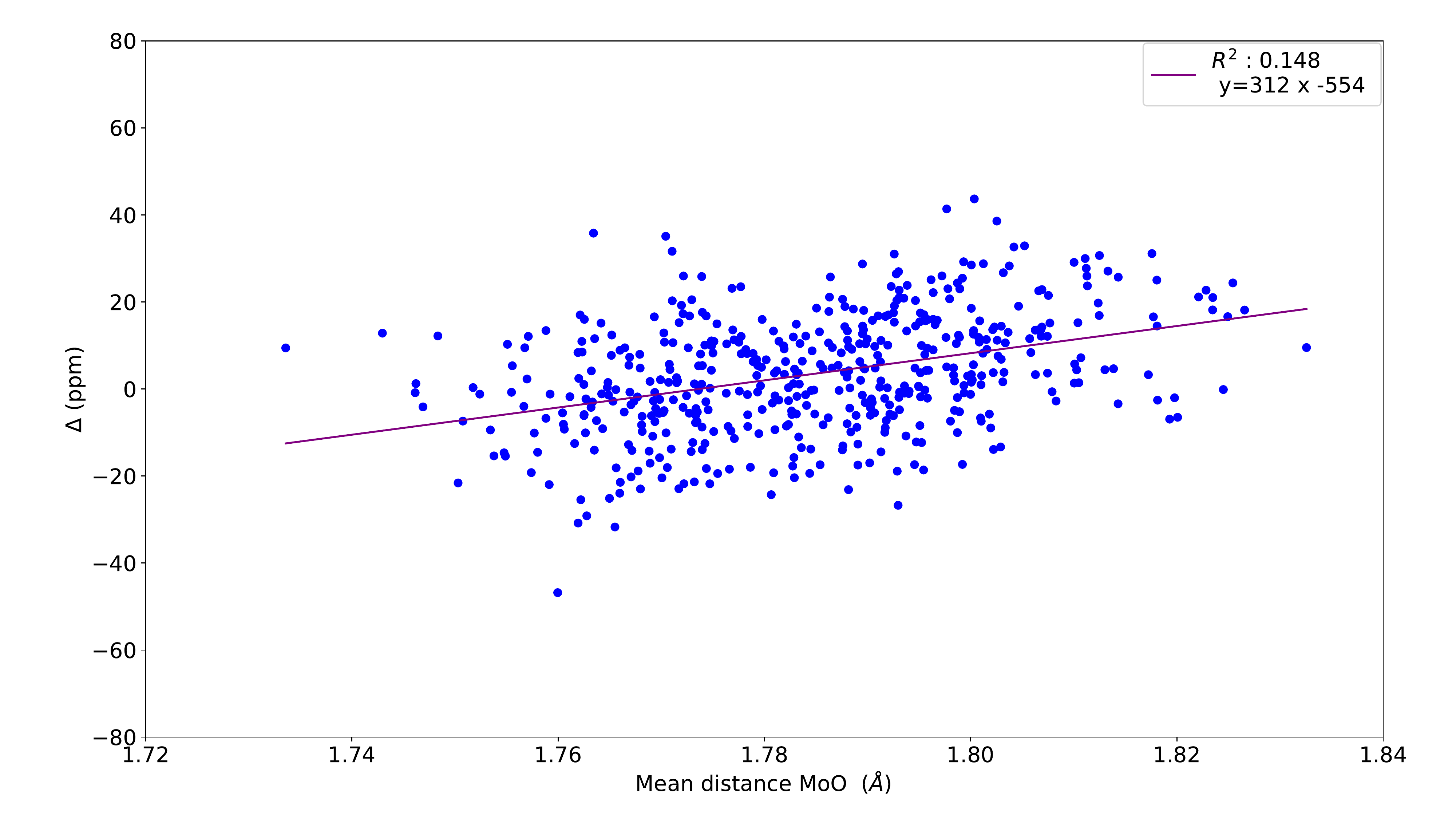}
	   \caption*{(b)}
	   \end{minipage}
	\begin{minipage}{0.90\linewidth}\centering
	   \includegraphics[width=1\linewidth]{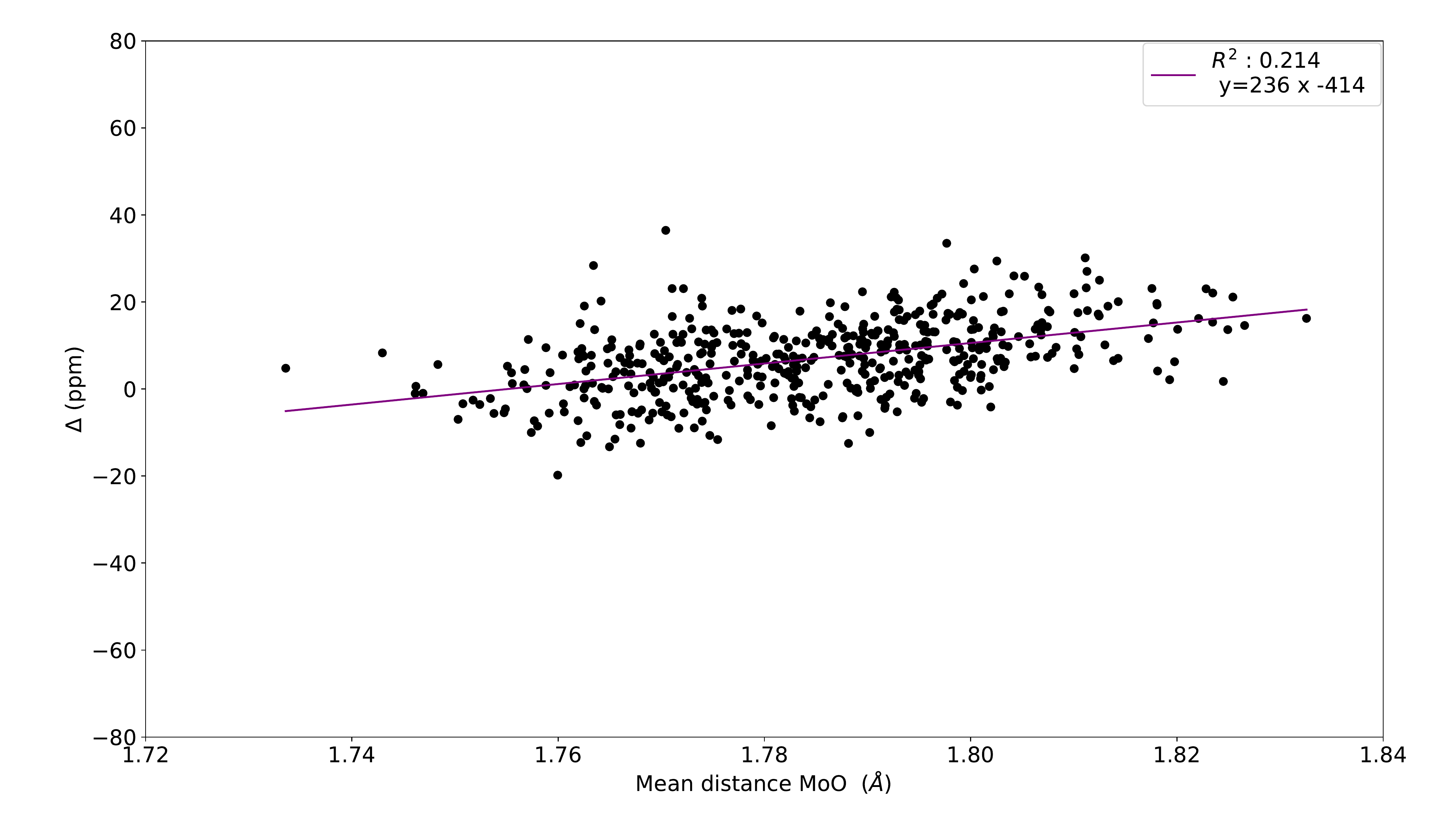}
	   \caption*{(c)}
	   \end{minipage}
	\caption{Difference between the embedded models (a) isolated $\ISO$; (b) FDE $\FDE$; and (c) FnT $\FnT$, and the supermolecule SR-ZORA results ($\Delta$, in ppm) as a function of the mean distance between Mo and the oxygen atoms in the Molybdate ion ($\bar{r}$[Mo-O]).}
	\label{fig:Error_ZORASR_vs_meandist}
\end{figure}

Since there is a non-negligible difference between the description of the supermolecular and embedded schemes, we have investigated the correlation between $\bar{r}$[Mo-O] and the values of the differences of $\sigma_{iso}$ obtained from embedding calculations and the supermolecular ones, denoted by $\Delta$ in Figure~\ref{fig:Error_ZORASR_vs_meandist}. From these results, we see that there is in fact no correlation between the two variables ($R^2 \leq 0.214$). {We also observe that there is a much larger dispersion of the data for the mechanical embedding than for the electronic embedding}. 

These $\Delta$ values are a measure of the error of a given embedding approximation with respect to the supermolecular standard, which therefore cannot be attributed to the molecular geometry represented by the $\bar{r}$[Mo-O] descriptor. It remains to be tested whether this error can be associated with other structural descriptors (for instance, ones taking into account the asymmetry around the Mo nucleus in a given environment) or whether it should be attributed to more subtle effects, for instance due to spin-orbit coupling, which we discuss in the following.

\subsubsection{Assessment of spin-orbit Hamiltonians}

\begin{table}[h]
\centering
\begin{tabular}{
        llc
}
\hline
{Model} & {$\bar{\sigma}_{iso} \pm std$} & {$\Delta\bar{\sigma}_{iso}$ wrt. $\SUP$}\\
\hline
{$ \SUP $} & $-585.0 \pm109.3$ & {-} \\
{$ \FnT $} & $-581.9 \pm104.2$ & $3.1 \pm8.8$ \\
{$ \FDE $} & $-585.8 \pm102.1$ & $-0.8 \pm13.7$ \\
{$ \ISO $} & $-600.3 \pm99.1$ & $-15.3 \pm24.7$ \\
\hline
\end{tabular}
\caption{Mean values for the isotropic NMR shielding constant ($\overline{\sigma}_{iso}$, in ppm) for the different computational models, and the respective standard deviations ($std$), obtained from the complete set of CPMD snapshots, employing the SO-ZORA Hamiltonian. $\Delta\bar{\sigma}_{iso}$ are the deviations with respect to $\SUP$.}
\label{tab:table_all_results_ZORASO}
\end{table}

In order to assess the importance of spin-orbit interactions on the the solvent effects, we begin by considering the SO-ZORA Hamiltonian. The 
$\bar{\sigma}_{iso}$ values and standard deviations for supermolecular and embedded calculations are shown in Table~\ref{tab:table_all_results_ZORASO}. 

From these we observe that there are no major differences with respect to the SR-ZORA results (see  Table~\ref{tab:table_all_results_ZORASR}) if it comes to the magnitude of the solvent effects for all approaches. We do observe some differences in the error patterns {(more readily grasped by inspecting the values of $\Delta\bar{\sigma}_{iso}$)}, with both electronic embedded models now being closer to the supermolecular $\bar{\sigma}_{iso}$ values than the mechanical embedding case, though once more the FDE $\bar{\sigma}_{iso}$ is the closest to the supermolecular one. The standard deviations in each embedded model show the same pattern as in the SR-ZORA, with FnT yielding the smallest and mechanical embedding the largest $std$ values.  

Comparing the results in Tables~\ref{tab:table_all_results_ZORASR} and~\ref{tab:table_all_results_ZORASO} we also observe significant changes on the absolute $\bar{\sigma}_{iso}$ values, which for SO-ZORA are upshifted by around 260 ppm from the SR-ZORA values. Such a spin-orbit effect on the absolute NMR shielding values is well-known and extensively discussed in the literature~\cite{lantto-jcp-125-184113-2006,jaszunski-handbook-compchem-497-2017}, and therefore we refer the readers to the review~\cite{jaszunski-handbook-compchem-497-2017} and recent examples~\cite{lantto-jcp-125-184113-2006,alkan-ssnmr-2018,ootani-jcp-125-2006,semenov-jpca-123-4908-2019,holmes-jctc-15-1785-2019}.

The nearly identical behavior of embedded SR-ZORA and SO-ZORA can be seen, first, as a manifestation of the fact that, since there are no significant spin-orbit coupling effects on the electronic structure of the solvent, the environment electron densities and electrostatic potentials will be rather similar in the two sets of calculations (SR and SO), as we can see from the rather similar standard deviations for all models considered. Second, since the Mo atom is not very heavy, spin-orbit coupling is not expected to bring about qualitative differences on the electron density of the \ce{MoO4^{2-}} ion. 

\begin{table}[h]
\centering
\begin{tabular}{
        ll
}
\hline
{Model} & {$\bar{\sigma}_{iso} \pm std$} \\
\hline
{$ \SUP ^ {216} $} & $-845.7 \pm 45.4$ \\ 
{$ \FnT ^ {216} $} & $-839.1 \pm 45.2$ \\
{$ \DIRAC $} & $-428.3 \pm 45.5$ \\
\hline
\end{tabular}
\caption{Mean values for the isotropic NMR shielding constant ($\overline{\sigma}_{iso}$, in ppm) for the different computational models, and the respective standard deviations ($std$), obtained from a subset of 216 CPMD snapshots chosen following the analysis of the normal probability plot (see supplemental information for details). The 4-component embedding calculations ({$\DIRAC$}) are performed using embedding potentials and frozen densities obtained from the SR-ZORA $\FnT^{216}$ model.}
	\label{tab:table_all_results_Dirac}
\end{table}

An appealing feature of the SR-ZORA and SO-ZORA calculations is their 
relatively modest computational cost {(respectively, around 3 and 8 
hours CPU time per snapshot to calculate all the three embedded models and the supermolecular system, with the latter taking about one order of magnitude longer than each of the embedding calculations)}~which has allowed us to consider a rather large number of 
snapshots. This makes them ideal 
tools for exploring the importance of relativistic effects across the periodic 
table while considering a great number of configuration from molecular 
dynamics or the effect of increasing the size of the active space. As 
discussed above, this allowed us to verify that SR-ZORA calculations already 
capture the solvent effects for the molybdate case. 

The exploration of a large number of snapshots also provided data with 
which to guide us in reducing the number of CPMD snapshots used for the 
more expensive calculations employing the DC Hamiltonian 
 {(about forty times more expensive than the ZORA-SO calculations in the current example)}.
With the help of normal probability plots (see supplemental information), we have been able to use the SR-ZORA data to reduce the number of snapshots to 216, roughly half of the original sample space, while at the same time assuring to keep a normal distribution of $\sigma_{iso}$ values for the SR-ZORA (supermolecule and embedded) calculations. We subsequently used this reduced sample to perform the DC calculations, for which results are shown in	Table~\ref{tab:table_all_results_Dirac}.

While we cannot directly compare the standard deviations from the DC 
results to those of Tables~\ref{tab:table_all_results_ZORASR} 
and~\ref{tab:table_all_results_ZORASO} we see from the table that the DC 
standard deviation values are close to the SR-ZORA ones, calculated with the 
same reduced sample of snapshots, and therefore we expect to see a similar 
error behavior for DC as for ZORA, had we used the original sample size. 
Furthermore, as was the case when improving the ZORA Hamiltonian with 
the addition of spin-orbit coupling, the more accurate treatment of 
relativistic effects afforded by the DC Hamiltonian introduces a further 
upshift of about 150 ppm with respect to SO-ZORA {mean 
difference}: $\bar{\sigma}_{iso}$, so that the  
difference {of the means} between the SR-ZORA and DC 
results is now a little over 410 ppm. 

This is a significant difference, given the growing interest in recent years in determining absolute shielding scales. Taking into account the steep computational cost of performing 4-component supermolecular calculations over hundreds of structures from MD simulations, we consider electronic embedding to be a promising way to calculate absolute NMR shielding constants in complex environments. 
 
\subsection{Larger active subsystems}

The results above suggest that the minimal structural model for solvated molybdate (comprising only the anion) is a rather good one for obtaining NMR shieldings in solution. In this section, we explore the effect of explicitly including nearest neighbor water molecules in the calculation. 

As we have found little quantitative differences in the behavior of the embedding approaches for the different Hamiltonians, and that it is computationally very expensive to perform a systematic expansion of the active subsystem while averaging over hundreds of snapshot, here we have decided to focus on selected SR-ZORA calculations. 

\begin{figure}
	\includegraphics[width=\linewidth]{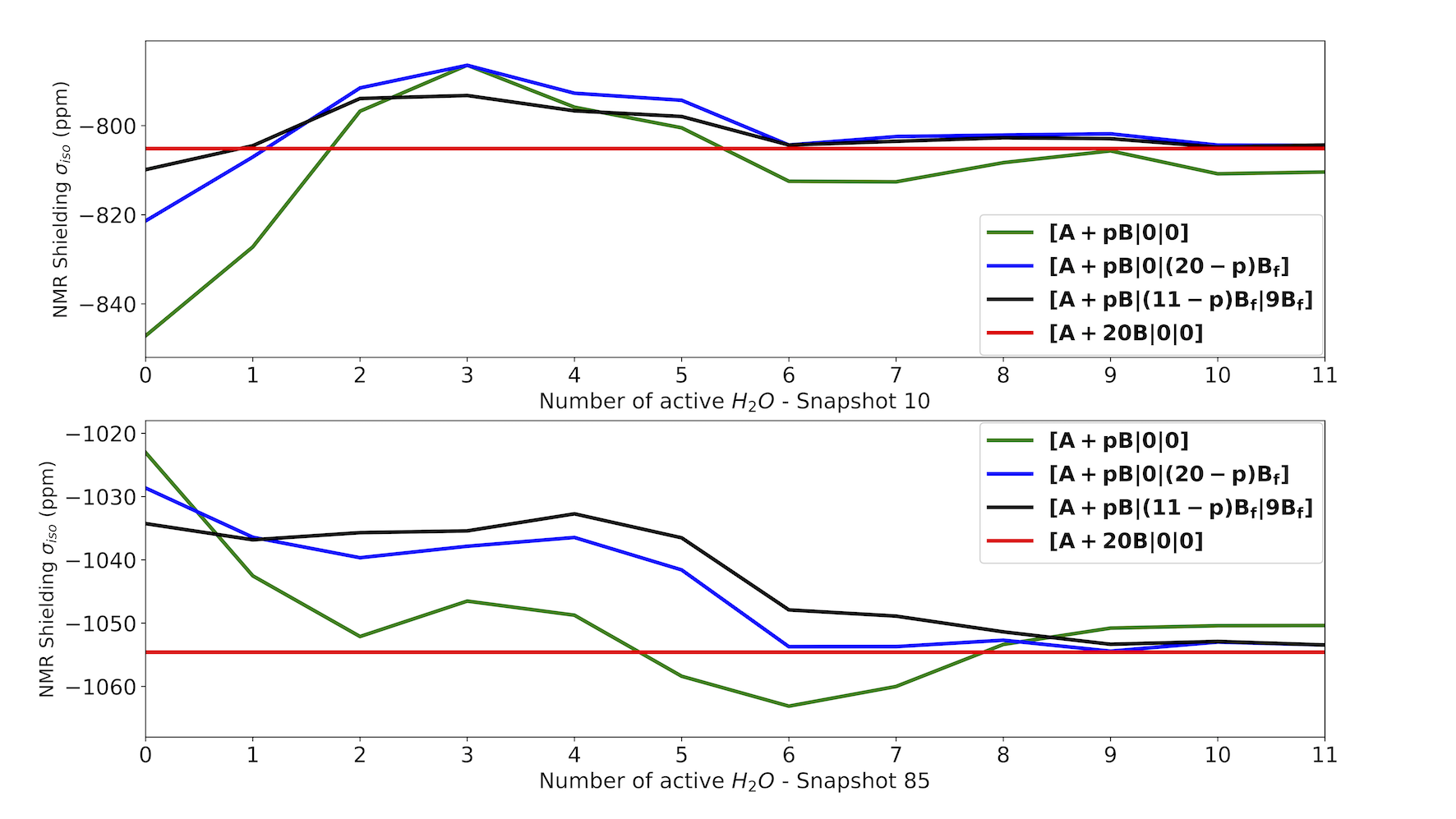}
	\caption{Variations of $\sigma_{iso}$ (in ppm) with increasing number of water molecules in the active subsystem for selected structures (top: snapshot 10; bottom: snapshot 85). All values are computed with the SR-ZORA Hamiltonian.}
	\label{graph:snap10_85_influence_N_eau}
\end{figure}

We therefore take two snapshots (numbered 10 and 85), representing respectively structures yielding NMR shieldings close to the $\overline{\sigma}_{iso}$ value and towards the  tails of the distribution. Figure~\ref{graph:snap10_85_influence_N_eau}{, in which we compare embedding models with increasing number of water molecules in the active subsystem,} summarizes our results. We only show results for up to 11 water molecules added to the active subsystem, due to the fact that no significant qualitative changes occur beyond this number. 

{Starting with now waters in the active subsystem ($p=0$),} for snapshot 10 (${\sigma}_{iso}$ closer to $\overline{\sigma}_{iso}$) we observe significant differences between embedding methods, with the subsystem embedding model ([$A +p B | (11-p)B_f | 9 B_f$ ]) yielding results which are only a few ppm below the reference supermolecular calculations. The FDE model ([$A +p B | 0 | (20-p)B_f$])) introduces more significant errors (over 10 ppm underestimation with respect to supermolecule), which from prior findings we can attribute to the importance of relaxing the environment species around a highly charged active subsystem. The mechanical embedding model fares the worst (over 40~ppm underestimation to supermolecule). 

Adding one water molecule to the active subsystem {($p=1$)} greatly improves the results for all embedding models, and gets the electronic embedding ones in very close agreement to the reference. From two {($p=2$)} to four {($p=4$)} explicit waters, however, we see a degradation of the electronic embedding results, which now overestimate ${\sigma}_{iso}$. At around six explicit water molecules {($p=6$)}, which roughly corresponds to the first solvation shell around the molybdate ion~\cite{Nguyen2015}, the electronic embedding methods have converged to the reference result, and show no further significant variations. The mechanical embedding model, on the other hand, shows small errors but does not show converged results even after 11 explicit water molecules {($p=11$)}. 

We note that a similar analysis has been carried out for  4-component calculations for snapshot 10, but restricting the reference calculations to 6 water molecules due to constraints in our computational resources (see computational details and supplemental information). As we observe the same trends in the DC calculations as in the SR-ZORA ones, we do not discuss the former explicitly. 

For snapshot 85 (${\sigma}_{iso}$ away from $\overline{\sigma}_{iso}$) we also observe the embedding approaches are pretty much converged to the reference after the first solvation shell is explicitly included. However, the behavior is quite different from snapshot 10 for the smaller number of explicitly included waters. Without any water molecule in the active subsystem, all methods underestimate ${\sigma}_{iso}$, and the mechanical embedding model is the worst among the three, but as the number of explicitly included water molecules increases, the electronic embedding methods do not show any improvement until that number reaches five.

The analysis of the snapshots in Figure~\ref{graph:snap10_85_influence_N_eau} provides a first, but somewhat indirect, indication that electronic embedding approaches are more reliable than mechanical embedding for two key structural models: the embedded \ce{MoO4^{2-}}, and the \ce{[MoO4(H2O)6]^{2-}} species. For a partial first solvation shell, it is nevertheless difficult to comprehend what is actually taking place from just following the ${\sigma}_{iso}$ values.

\subsection{Real-space analysis of embedding\label{analysis-shielding}}

One can try to identify the underlying differences between embedding and supermolecular calculations by investigating the systems in real space. One option is to  follow the differences in the electron density for the two treatments, as often done for other properties and previously done by Bulo and coworkers~\cite{bulo-jpca-112-2640-2008} for NMR shieldings. {For convenience in the analysis, we decided to deviate slightly from the models discussed above and collected a number of water molecules closest to the molybdate ion into a single subsystem, while any molecules further away are still considered as individual subsystems. While such a modification appears to have little effect on the calculated shieldings (see Table~\ref{tab:val_snap10_11grouped} in the supplemental information), we discuss below its effect on density.} 

\begin{figure}
\centering
	\begin{minipage}{0.48\linewidth}\centering
	   \includegraphics[width=0.9\linewidth]{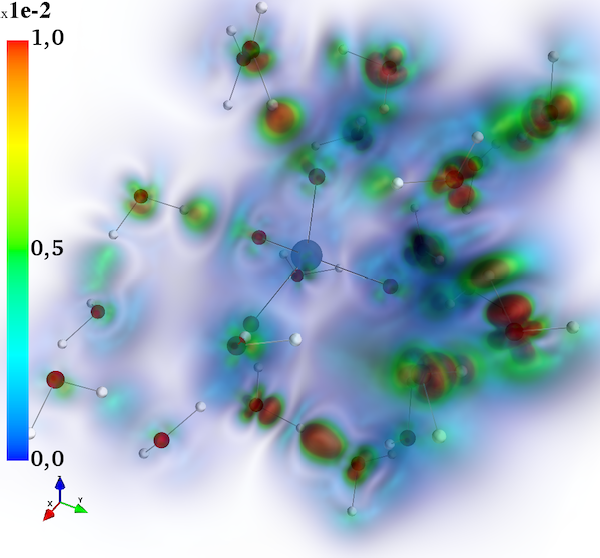}
	     \caption*{(a)}
	   \end{minipage}
		\begin{minipage}{0.48\linewidth}\centering
	   \includegraphics[width=0.9\linewidth]{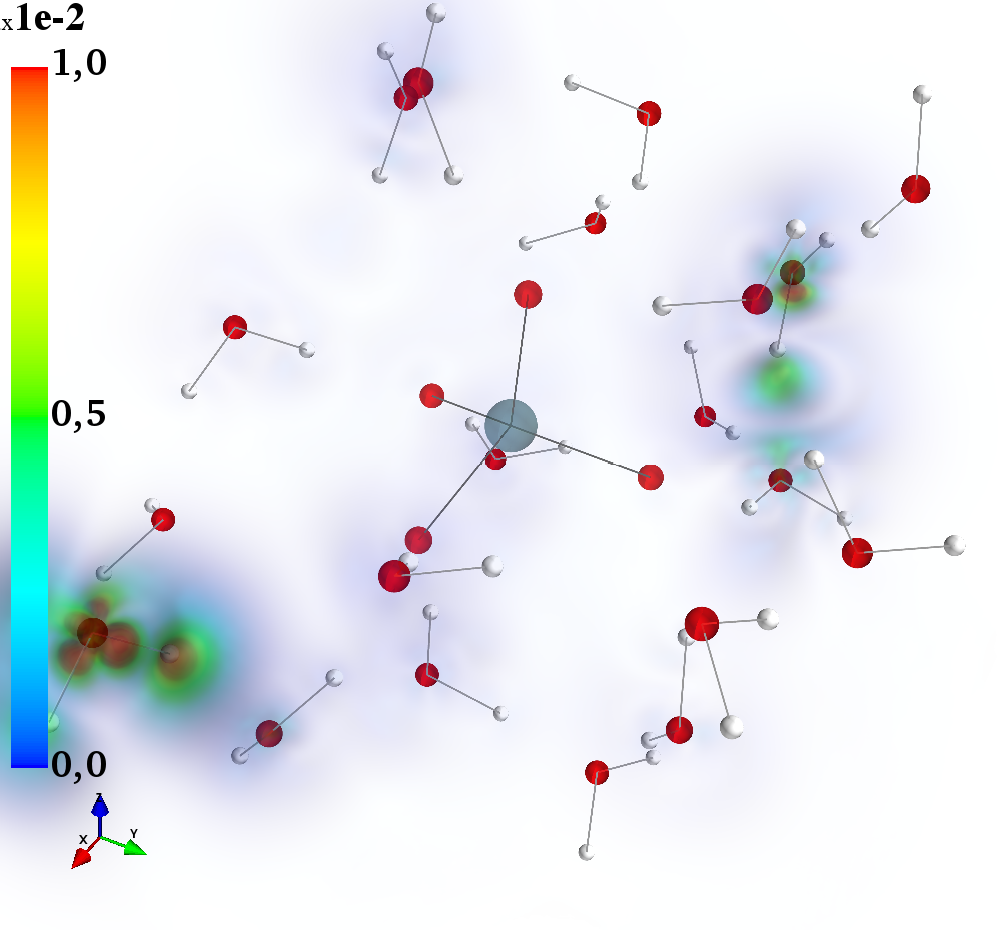}
	     \caption*{(b)}
	   \end{minipage}
	\caption{Volumetric plot of the difference of SR-ZORA electron densities between (a) $[A|11B_g|9B_f]$ and $\SUP$; and (b) $\FnT$ and $[A|11B_g|9B_f]$ for snapshot 10. Displayed are difference density values at each grid point between 0.00 and 0.01 a.u.} 
	\label{graph:densite_2_B}
\end{figure}

Figure~\ref{graph:densite_2_B} presents the difference in SR-ZORA density between the reference and the electronic embedding models ($\FnT$ and $\FDE$). It is important to note here that this figure does not present density isosurfaces, but rather a volumetric description (i.e.\, the accumulated values ranging from lower (0.00) to the upper (0.01) bound considered) of the electron density. This representation has the advantage of enabling the visualization of high and low density regions in the same plot.

From Figure~\ref{graph:densite_2_B} we clearly see that inaccuracies in the 
embedding calculations, due to (for instance) the approximate treatment of 
the non-additive kinetic energy term, introduce small errors in the 
densities throughout the whole space. These errors tend to build up (green 
to red colors) in the regions away from the core of the \ce{MoO4^{2-}} unit, 
and for the $\FnT$ model they tend to be more significant around the 
second solvation shell, given that first solvation shell is fully relaxed in the 
presence of \ce{MoO4^{2-}} 
 {(Figure~\ref{graph:densite_2_B}-a)}. We 
also see from that figure that there is a small but non-negligible effect on 
the densities of the environment when we consider the molecules as a 
single subsystem or as a collection of fragments--as 
 {(Figure~\ref{graph:densite_2_B}-b)}, in the latter, we create 
additional frontiers between subsystems, due to the monomer expansion 
and the eventual buildup of errors due to the approximate kinetic energy 
functionals.
 
Whatever the case, the analysis of the electron density is at odds with our findings for the mean values of shielding, since there appears to be nothing that points to significant errors in the region of the Mo-O bonds. This suggests that trying to understand the behavior ${\sigma}_{iso}$ values from changes in the electron density is not a suitable strategy.

An alternative to the visualization of the electron density is the analysis of the property densities. We have provided a first example for such an analysis for the analysis of embedding in the description of our 4-component implementation of subsystem embedding theory for magnetic properties, but for rather small systems~\cite{Olejniczak2017}. Here we provide the analysis of a more complex example, while at the same time forsaking the use of isosurfaces in favor of volumetric plots.

\begin{figure}[htp]
	 p = 2\\
	\begin{minipage}[t]{0.33\linewidth}
		\includegraphics[width=\linewidth]{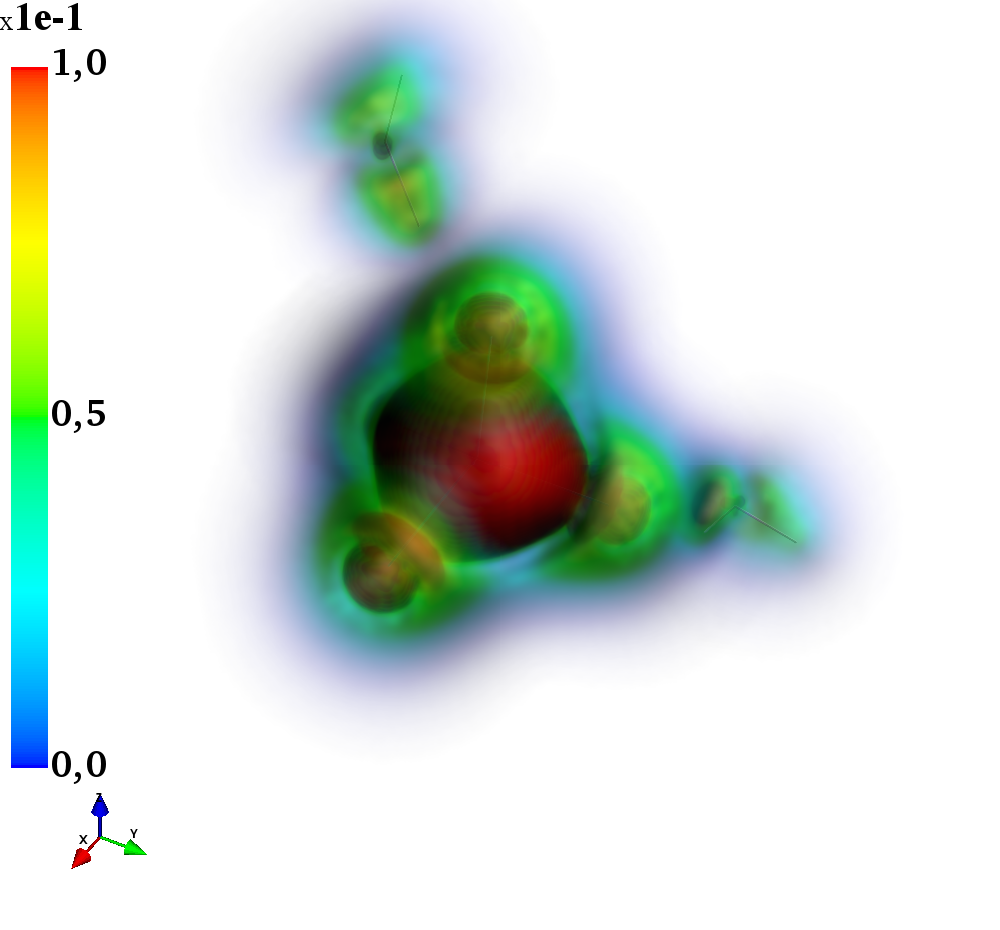}
		\label{graph:6a_p2}
	\end{minipage}\hfill
	\begin{minipage}[t]{0.33\linewidth}
		\includegraphics[width=\linewidth]{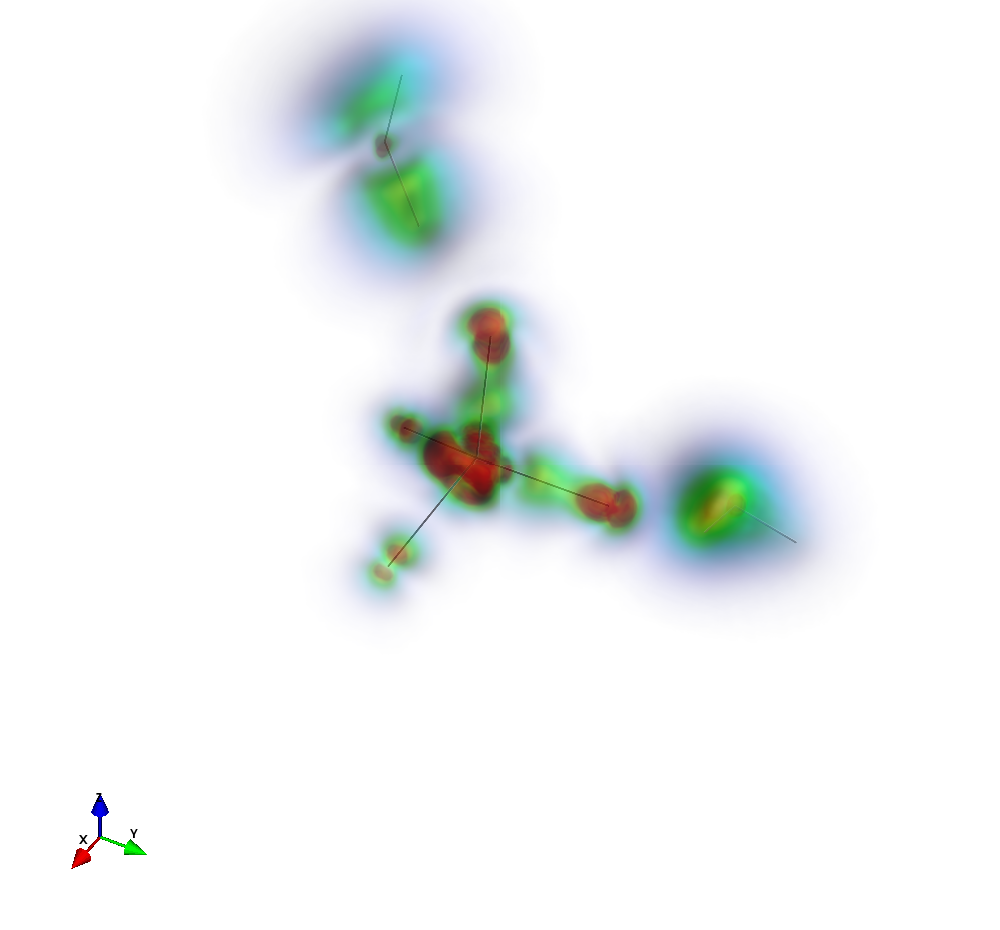}
		\label{graph:6b_p2}
	\end{minipage}\hfill
	\begin{minipage}[t]{0.33\linewidth}
		\includegraphics[width=\linewidth]{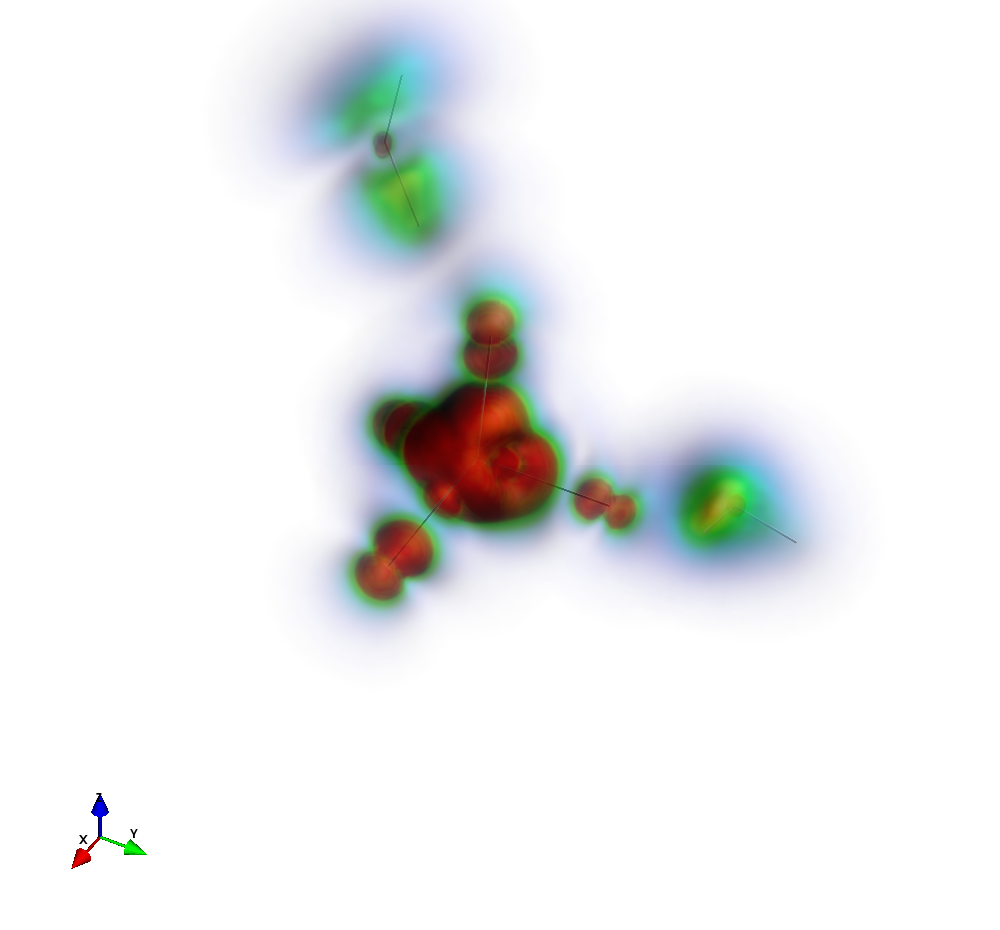}
		\label{graph:6c_p2}
	\end{minipage}\hfill
	\vspace{-1\baselineskip}
	p = 4\\	%
	\begin{minipage}[t]{0.33\linewidth}
		\includegraphics[width=\linewidth]{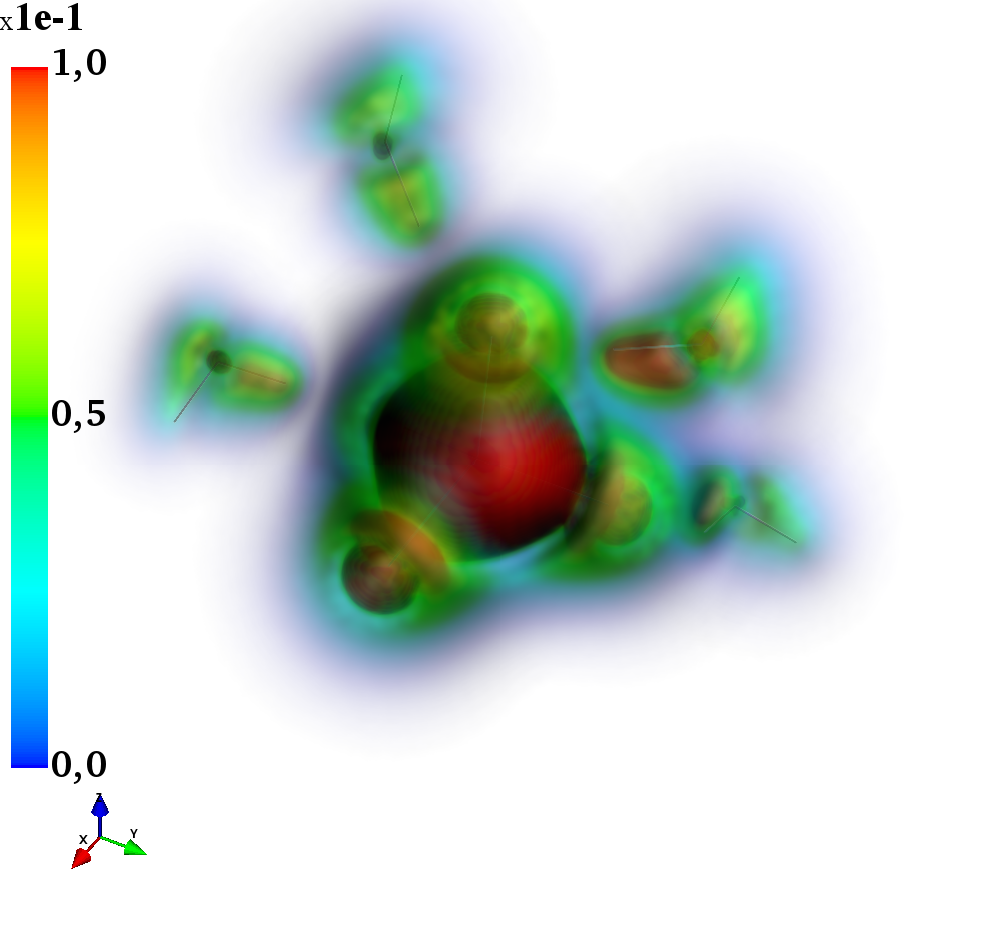}
		\label{graph:6a_p4}
	\end{minipage}\hfill
	\begin{minipage}[t]{0.33\linewidth}
		\includegraphics[width=\linewidth]{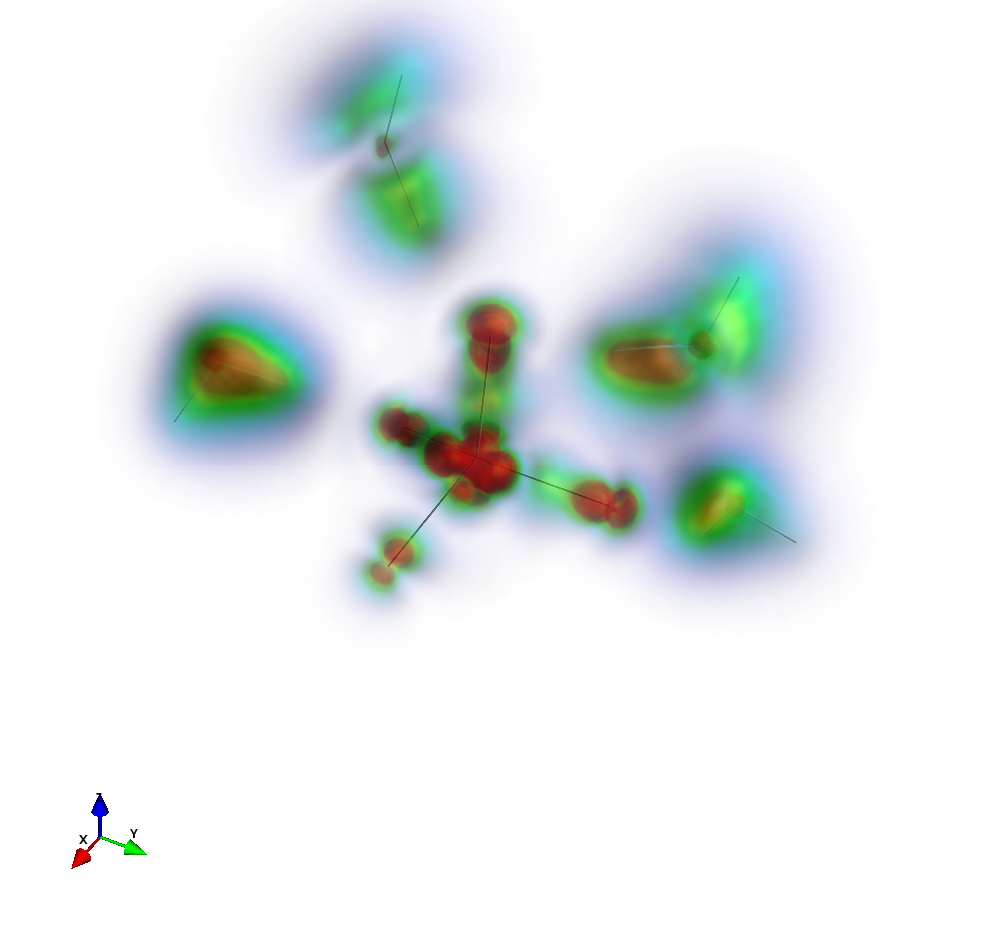}
		\label{graph:6b_p4}
	\end{minipage}\hfill
	\begin{minipage}[t]{0.33\linewidth}
		\includegraphics[width=\linewidth]{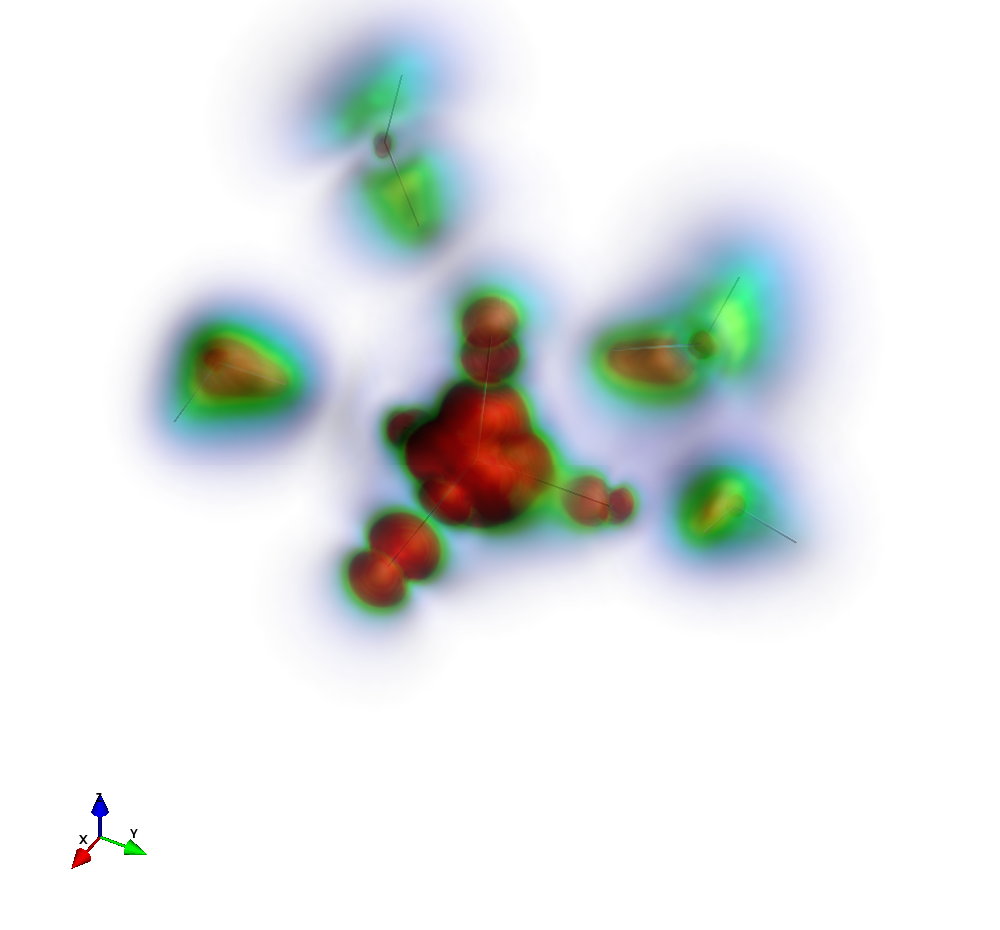}
		\label{graph:6c_p4}
	\end{minipage}\hfill 
	\vspace{-1\baselineskip}
	 p = 6\\%
	\begin{minipage}[t]{0.33\linewidth}
		\includegraphics[width=\linewidth]{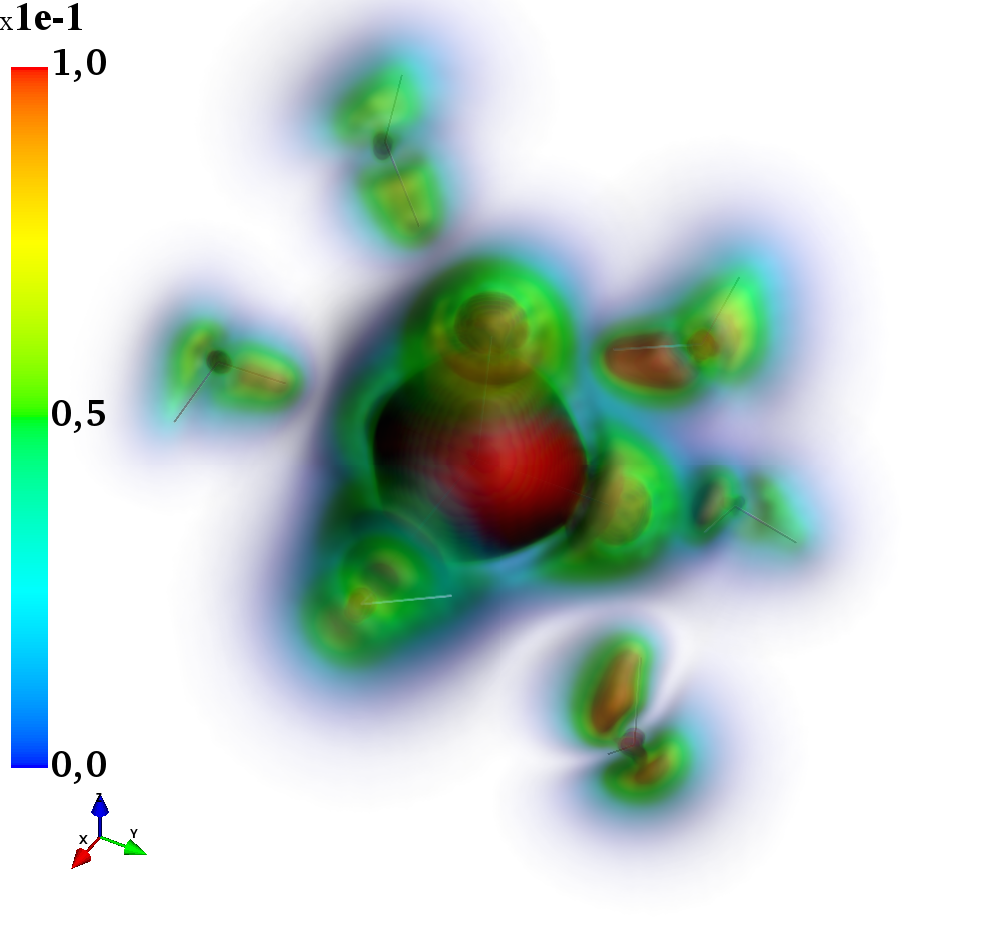}
		\label{graph:6a_p6}
	\end{minipage}\hfill
	\begin{minipage}[t]{0.33\linewidth}
		\includegraphics[width=\linewidth]{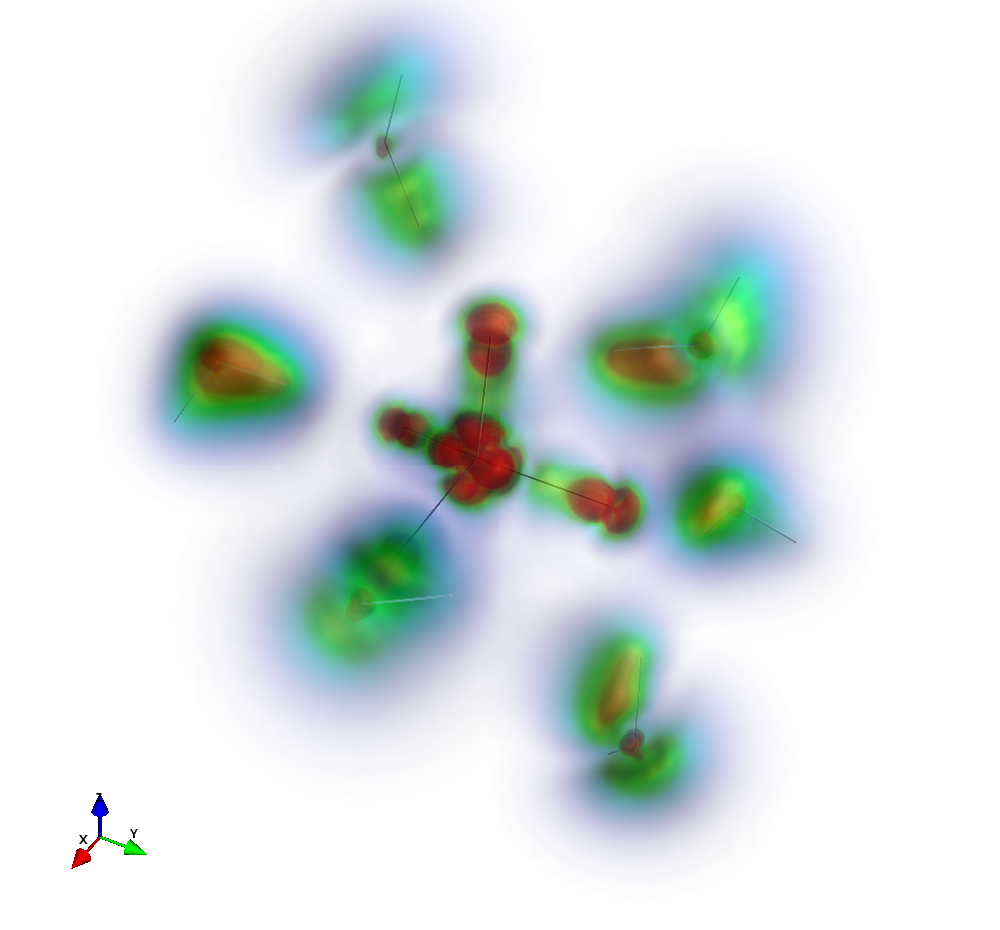}
		\label{graph:6b_p6}
	\end{minipage}\hfill
	\begin{minipage}[t]{0.33\linewidth}
		\includegraphics[width=\linewidth]{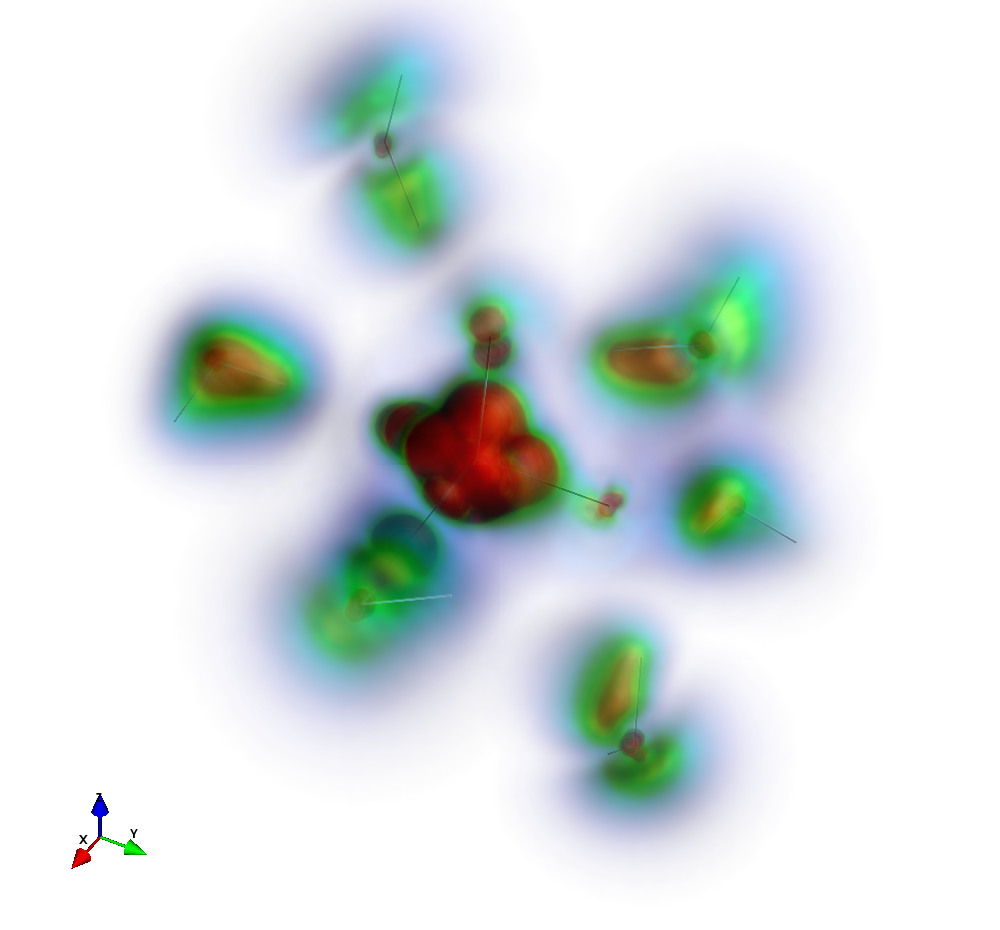}
		\label{graph:6c_p6}
	\end{minipage}\hfill %
	\caption{Volumetric plots of shielding densities and differences in 
	shielding densities between 0.0 and 0.1 ppm, computed with the DC 
	Hamiltonian, for \ce{MoO4(H2O)_p]^{2-}} with increasing 
	number of $p$ water molecules. Left column: Supermolecule 
	($[A+pB|0|0]$); Middle column: difference between supermolecular and 
	electronic embedding ($[A|6B_g|0]$) calculations; Right 
	column:difference between supermolecular and mechanical embedding 
	($[A|0|0]+[0|6B_g|0]$) calculations. For embedded models, water 
	molecule contributions to \ce{^{95}Mo} shielding densities are calculated 
	with the NICS method.}
	\label{graph:densite_shielding_SUP}
\end{figure}

Because the analysis of shielding densities is only implemented in the DIRAC code, we have restricted ourselves to the \ce{[MoO4(H2O)_p]^{2-}} species ($p = 1-6$). These species are then considered as our references, and both mechanical and electronic embedding calculations considering only the molybdate ion in the active subsystem have been performed for each value of $p$. All fragments have therefore been treated with the DC Hamiltonian, with the embedding potentials being determined with \emph{freeze-thaw} calculations using DIRAC. The results of these calculations are found in  Figure~\ref{graph:densite_shielding_SUP}, where we present the  shielding density for the \ce{^{95}Mo} atom on the references, along with the differences in shielding density between the references and the electronic and mechanical embedding results. For the embedding calculations, contributions from the solvent water molecules at the location of the Mo atom are obtained with the NICS procedure. 

The first striking feature of the embedding results is that, unlike for the electron densities, the error in the embedding calculations is indeed localized within the \ce{MoO4^{-2}} species, in particular around the Mo atom. In addition to that, there are smaller errors also at the positions of the molybdate oxygen atoms. 

We believe this points to a physical process that, while involving a rather local operator (the hyperfine operator has an effective $r^{-2}$ dependence), makes the resulting property ${\sigma}_{iso}$ (described as the cross product of the hyperfine and the magnetically induced current density) have a much less localized nature, and with that non-negligible contributions rather far from the responding atom (in the vicinity of the oxygen atoms) arise. 

This goes to explain why, in both types of embedding, the dependence of ${\sigma}_{iso}$ on the number of water molecules is so significant: while the water molecules themselves do not contribute significantly to the shielding density, their absence (in the case of mechanical embedding) or the relatively inaccurate representation of the active and environment subsystems (in the case of electronic embedding) is enough to perturb the contribution to the shielding density around the molybdate oxygen atoms.

Furthermore, we see that as the first solvation shell is built up by including the nearest water molecules to the active subsystem (we show in Figure~\ref{graph:densite_shielding_SUP} only the even $p$ values, see supplemental information for the odd values), the perturbation on the molybdate oxygen atoms is not accounted for in a systematic manner by the embedding methods, so that errors may build up even in regions in which water molecules had already been added. We do not yet possess the analysis tools to fully understand the interplay between these different effects, and are currently pursuing the development of new analysis approaches to address the issue.

Beyond these similarities, we nevertheless see that the errors for the mechanical embedding calculations, though of similar magnitude than those for the electronic embedding, extend much farther than for the latter, and explain why electronic embedding is more reliable.

\section{Conclusions and perspectives\label{conclusions}}

In this manuscript we provide an assessment of approaches based upon the 
frozen density embedding (FDE) framework for the creation of 
computationally efficient models capable of capturing solvent effects on 
NMR shieldings of heavy elements. We have investigated how water 
influences the shielding of \ce{^{95}Mo} in the molybdate dianion 
(\ce{MoO4^{2-}}),  which can be at times a reference for \ce{^{95}Mo} NMR 
experiments, or a precursor for building Mo-containing materials.

A particular strength of FDE in this context is the ease with which different 
relativistic Hamiltonians can be used for different parts of the system. Here 
it allowed to compare the scalar relativistic (SR) and 2-component 
spin-orbit (SO) ZORA Hamiltonians, and the 4-component Dirac-Coulomb 
(DC) Hamiltonian to describe the active subsystem while using the SR ZORA 
Hamiltonian for the aqueous environment.

From our calculations on different structural models, employing a large set 
of structures obtained from CPMD trajectories from work previously 
described in the literature, we have established that for the \ce{^{95}Mo} 
mean isotropic shielding value, there are weak but non-negligible solvent 
shifts, due to the modification of the response of the electronic 
wavefunction of solute by the solvent. We find that even an approach along 
the lines of mechanical embedding yields a mean isotropic shielding value 
in good agreement with reference results obtained with standard DFT 
calculations on the full system.

Such mechanical embedding calculations, however, show a much larger spread of shielding values (for the different CPMD snapshots) with respect to the reference calculations than either of the electronic embedding variants considered. This leads us to conclude that electronic embedding is in fact an essential component for computational models.  

By investigating the dependence of the isotropic shielding value from embedding calculations on the size of the active subsystems for selected CPMD snapshots, we have observed that convergence to the reference calculations it not monotonic and large discrepancies may still occur until roughly a full first solvation shell is explicitly included in the active subsystem~\cite{fukal-pccp-21-9924-2019}.

We have found our results to be stable with respect to changes in the way the environment was described by FDE, that is, irrespective of whether we used embedding potentials obtained with the same (SR-ZORA or SO-ZORA) Hamiltonian throughout, or by using a combination of SR-ZORA based environment electron densities and electrostatic potentials together with a DC based calculation for the active subsystem. 

This points to the possibility of devising efficient computational models for Mo-containing compounds, in which  SR-ZORA DFT calculations are used to prepare embedding potentials for more sophisticated, 4-component based calculations on the Mo-containing regions of interest. The latter approach is particularly interesting to investigate the absolute shieldings for molecules in complex environments.

We have found a significant shift (around 260 ppm) when changing from the SR-ZORA to the SO-ZORA Hamiltonian, and another significant shift (around 150 ppm) when changing from the SO-ZORA to the DC Hamiltonians. However, given the rather systematic nature of the differences observed, for chemical shifts of Mo complexes it is likely that the SR-ZORA Hamiltonian provides sufficient accuracy.
 
Finally, we have explored the visualization of the electron and shielding densities, as a means to provide further insight on the mechanisms behind the solvent effects, in addition to the performance (and shortcomings) of the embedding approaches. Our results underscore the point we have made previously, in that changes in shielding densities are better understood from the analysis of the differences of the corresponding shielding densities rather than from the changes in electron density. 

Furthermore, we observe that the major source of errors in embedding calculations comes from the regions around the Mo nucleus, with less important discrepancies coming from all around the \ce{MoO4^{2-}} species. As the major difference between the FDE and reference calculations lies in the use of approximate kinetic energy functionals to describe the respective non-additive contributions, this suggests that for magnetic properties it would be of interest to try to improve the performance of such functionals.

\section{Acknowledgements}

The authors thank Dr.\ Jérôme Cuny (Laboratoire de Chimie et Physique Quantiques, UMR 5626, Toulouse) for kindly providing us with the snapshots from the CPMD simulations on the molybdate ion.

The members of the PhLAM laboratory acknowledge support from the CaPPA project (Chemical and Physical Properties of the Atmosphere), funded by the French National Research Agency (ANR) through the PIA (Programme d'Investissement d'Avenir) under contract ``ANR-11-LABX-0005-01'' and I-SITE ULNE project OVERSEE (ANR-16-IDEX-0004), as well as by the Ministry of Higher Education and Research, Hauts de France council and European Regional Development Fund (ERDF) through the Contrat de Projets Etat-Region (CPER) CLIMBIO (Changement climatique, dynamique de l'atmosph\`ere, impacts sur la biodiversit\'e et la sant\'e humaine), the CNRS Institute of Physics (INP) via the PICS program (grant 6386), and computational time provided by the French national supercomputing facilities (grants DARI A0030801859, A0050801859).

MO acknowledges support from the Polish National Science Centre (NCN) (grant number 2016/23/D/ST4/03217).

\section*{Conflict of interest}
There are no conflicts to declare.

\section*{Supporting Information Available:}

The supporting information is available as appendix on page~\pageref{sec:appendix}.

\bibliography{references.bib}

\clearpage
\onecolumngrid
\appendix
\renewcommand{\thefigure}{S\arabic{figure}}
\renewcommand{\thetable}{S\arabic{table}}
\setcounter{figure}{0}
\setcounter{table}{0}
\section{Supplemental information\label{sec:appendix}}
\subsection{Influence of the embedding model on \ce{^{95}Mo} shielding values}

\begin{table}[htbp]
\begin{tabular}{
        l
        S[round-mode=places,round-precision=1]
        S[round-mode=places,round-precision=1]
        c
        S[round-mode=places,round-precision=1]
        S[round-mode=places,round-precision=1]
}
\toprule
& \multicolumn{2}{c}{snapshot 10}&& \multicolumn{2}{c}{snapshot 85}\\
\cline{2-3}\cline{5-6}
{Model} & {$\sigma_{iso}$} & $\Delta\sigma_{iso}^{ref}$ && {$\sigma_{iso}$} & $\Delta\sigma_{iso}^{ref}$\\ 
\midrule
{$\SUP$ [$ref$]} 			& -805.141 & && -1054.585 & \\
{$\FnT$}			& -808.887 & -3.746 	&& -1034.50x7 & 20.078\\
{$[A|11B_g|9B_f]$}	& -808.879 & -3.738		&& -1035.921 & 18.664\\
{$\FDE$}			& -822.851 & -17.710 	&& -1028.881 & 25.704\\
{$[A|0|11B_g+9B_f]$}& -822.446 & -17.305	&& -1032.424 & 22.161\\  
\bottomrule
\end{tabular}
\captionof{table}{SR-ZORA values of $\sigma_{iso}$ as a function of the fragmentation of the first solvation layer of \ce{MoO4^{2-}}, and $\Delta\sigma_{iso}^{ref}$, with respect to the supermolecule $\SUP$ reference, for snapshots 10 and 85.}
\label{tab:val_snap10_11grouped}
\end{table}

The electron densities and electrostatic potential for the environment are obtained either from calculations on individual water molecules (and we refer to a ``fragmented'' environment, noted $f$) or by these forming a single subsystem (and we refer to a ``grouped'' environment, noted $g$).
We have looked at the influence of these two wait of treating the water molecules for two snapshots of the CPMD trajectories, snapshot 10 yielding results close to the average and snapshot 85 standing in the tail of the result distribution. 

Table~\ref{tab:val_snap10_11grouped} reveals that the $\sigma_{iso}$ values obtained with FDE and FnT embeddings are marginally sensitive to the  way the embedding density is calculated. For the sake of minimizing the computational cost, we will rely on ``fragmented'' water clusters for the generation of the FnT and FDE embedding potentials.

\subsection{Influence of the relativistic Hamiltonian on $\sigma_{iso}$}

\begin{figure}[htp]
	\includegraphics[width=15cm]{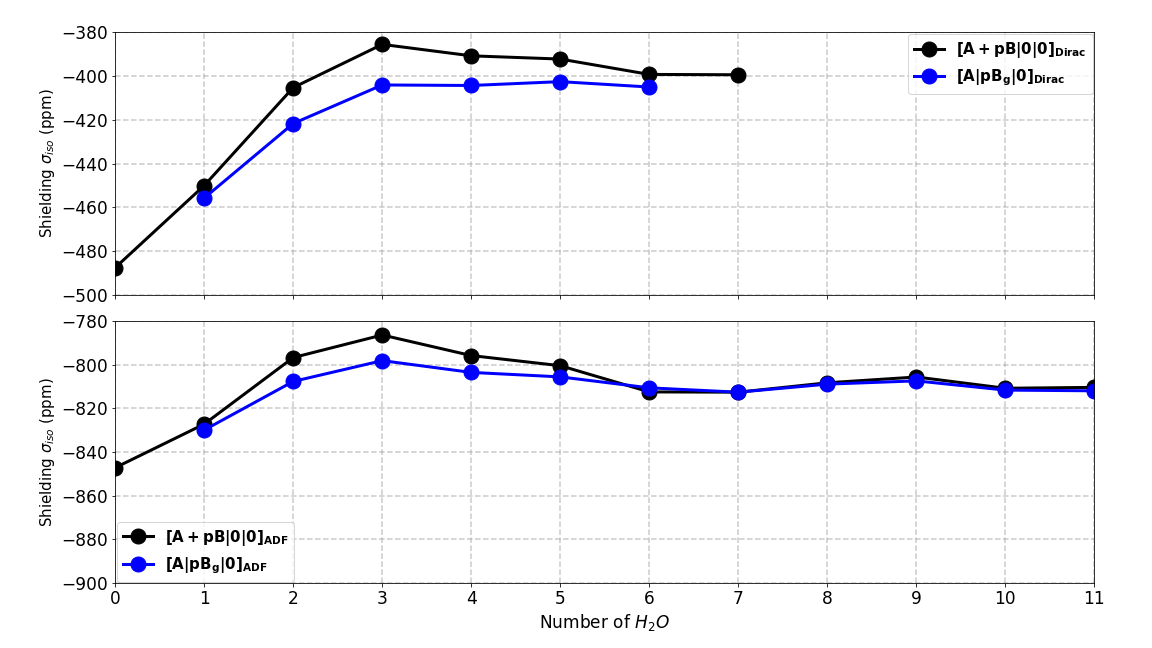}
	\caption{Evolution of the isotropic shielding $ \sigma_{iso} $ (in ppm) as 
	a function of the number of water molecules $ p $ in the active system ($ 
	[A+pB|0|0] $) or in the FnT process ($ [A|pB_g|0] $), with Dirac (top) and with 
	ADF (bottom), for snapshot 10.}
	\label{graph:All_snap10_bis}
\end{figure}

In the manuscript, we have noted that the change of relativistic Hamiltonian from SR-ZORA (ADF) to DC (DIRAC) translates into a change \SI{410}{ppm} of the absolute shielding constant for the $\FnT$ model (See Table~\ref{tab:table_all_results_Dirac}). To verify whether this shift is independent of the chemical model, we have performed a systematic comparison of $\sigma_{iso}$ values obtained with SR-ZORA and DC Hamiltonians for molybdate \ce{[MoO4(H2O)_p]^{2-}} hydrated by $p=1,6$ water molecules either treated explicitly ($[A+pB|0|0]_{ADF}$ and $[A+pB|0|0]_{Dirac}$), or by an embedding potential treating the \ce{(H2O)_p} water molecules as a ``grouped'' water cluster  ($[A|pB_g|0]_{ADF}$ and $[A|pB_g|0]_{Dirac}$. The values drawn on Figure~\ref{graph:All_snap10_bis} reveal very similar trends between the SR-ZORA and DC values. For both Hamiltonians, the difference between FnT ($[A|pB_g|0]$) values and the supermolecular $[A+pB|0|0]$), vanishes at $p=6$ water molecules.

\subsection{Statistical analysis of $\sigma_{iso}$ values; selection of a snapshot subset}
\begin{figure}[htp]
	\includegraphics[width=15cm]{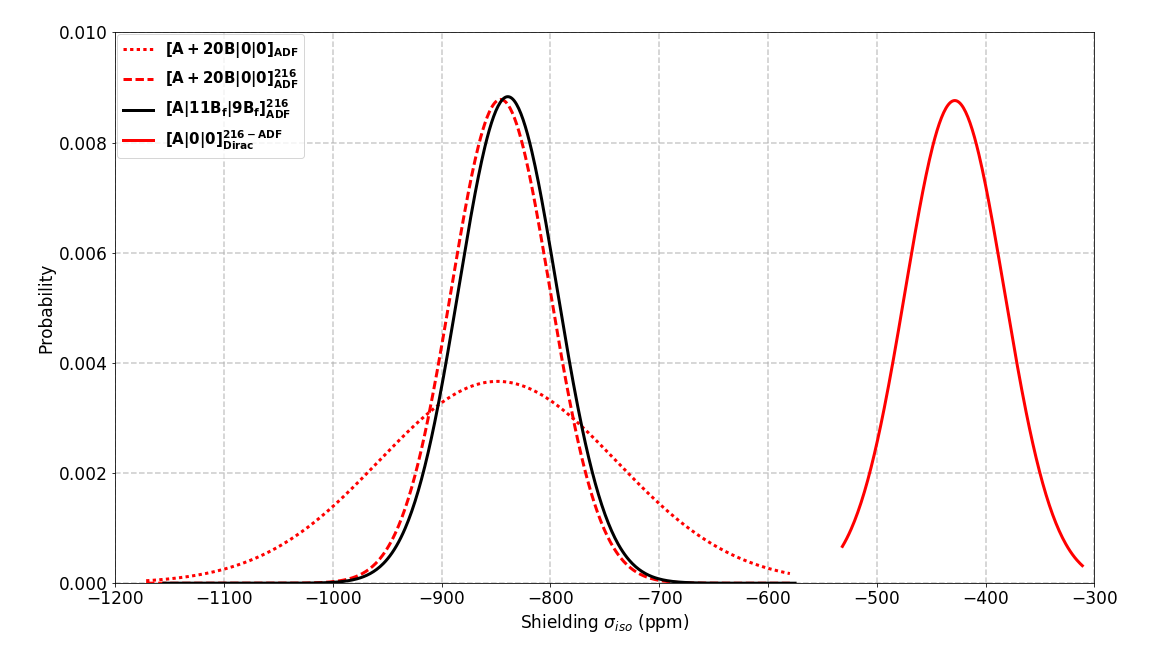}
	\caption{Normalized Gaussians representing $\sigma_{iso} $ (in ppm).
	The red dotted line refers to supermolecular ADF for all snapshots ($ [A+20B|0|0]_{ADF} $), while the red dashed line refers to the subset of 216 selected snapshots ($ [A+20B|0|0]^{216}_{ADF}$). The FnT ADF calculations ($ [A|11B_f|9B_f]^{216}_{ADF} $) for this 216 snapshot subset are drawn with a black line, while the DIRAC calculations using SR-ZORA (ADF) embedding potentials and electrostatic potentials are drawn in red.}
	\label{graph:All_ADF_DIRAC_b}
\end{figure}

\begin{figure}[htp]
	\includegraphics[width=15cm]{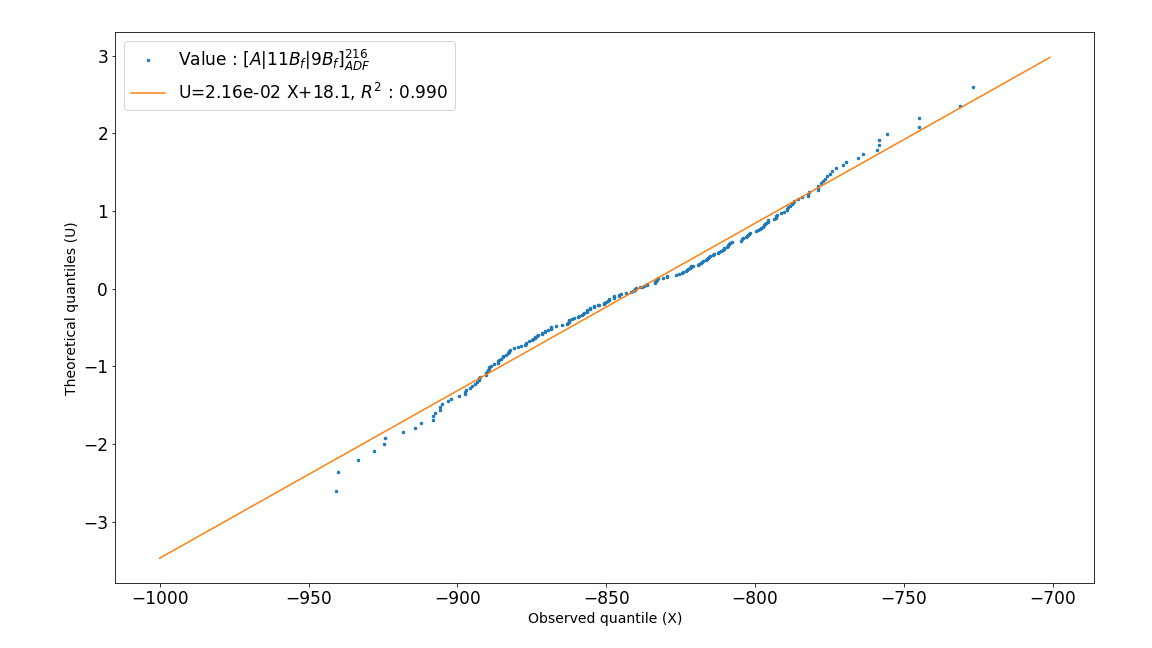}
	\caption{Normal probability plot of $ \sigma_{iso} $ for the 216 
	selected snapshots with ADF (FnT)~:~$ [A|11B_f|9B_f]^{216}_{ADF} $.} 
	\label{graph:Henry_line_ADF_FnT_216}
\end{figure}

\begin{figure}[htp]
	\includegraphics[width=15cm]{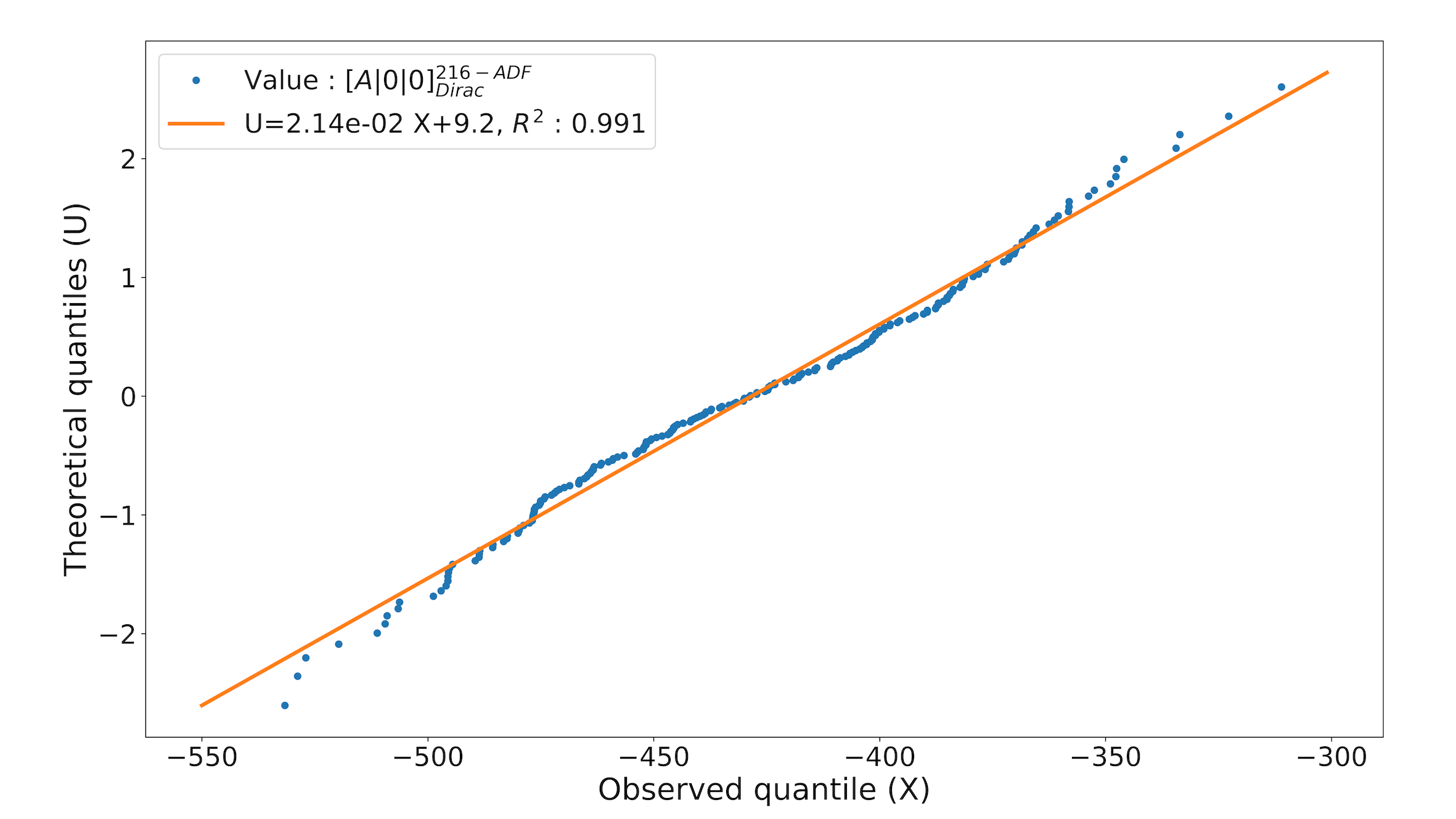}
	\caption{Normal probability plot of $ \sigma_{iso} $ for the 216 
	selected snapshots with $ \DIRAC $~:~$ [A|0|0]^{216-ADF}_{Dirac} $.} 
	\label{graph:Henry_line_Dirac_et_ADF_216}
\end{figure}

Figure~\ref{graph:All_ADF_DIRAC_b} plots the distribution of the $\sigma_{iso}$ values obtained from the 517~snapshots (red dotted line). With the aim of performing DC calculations, which are computationally more expensive that the ADF ones, it was necessary to reduce the number of snapshots, but keeping a normal distribution. The normal probability plots of Figures~\ref{graph:Henry_line_ADF_FnT_216} and~\ref{graph:Henry_line_Dirac_et_ADF_216}, reveal that the ADF values, both for supermolecular calculations (dashed red line) and FnT calculations (black line) obtained for a subset of 216~snapshots follow a normal law, though with a smaller variance. The DC values (red lines) have the same statistical distribution, shifted by \SI{411}{ppm}.

\begin{figure}[htp]
 p = 1 \\ 
	\begin{minipage}[t]{0.33\linewidth}
		\includegraphics[width=\linewidth]{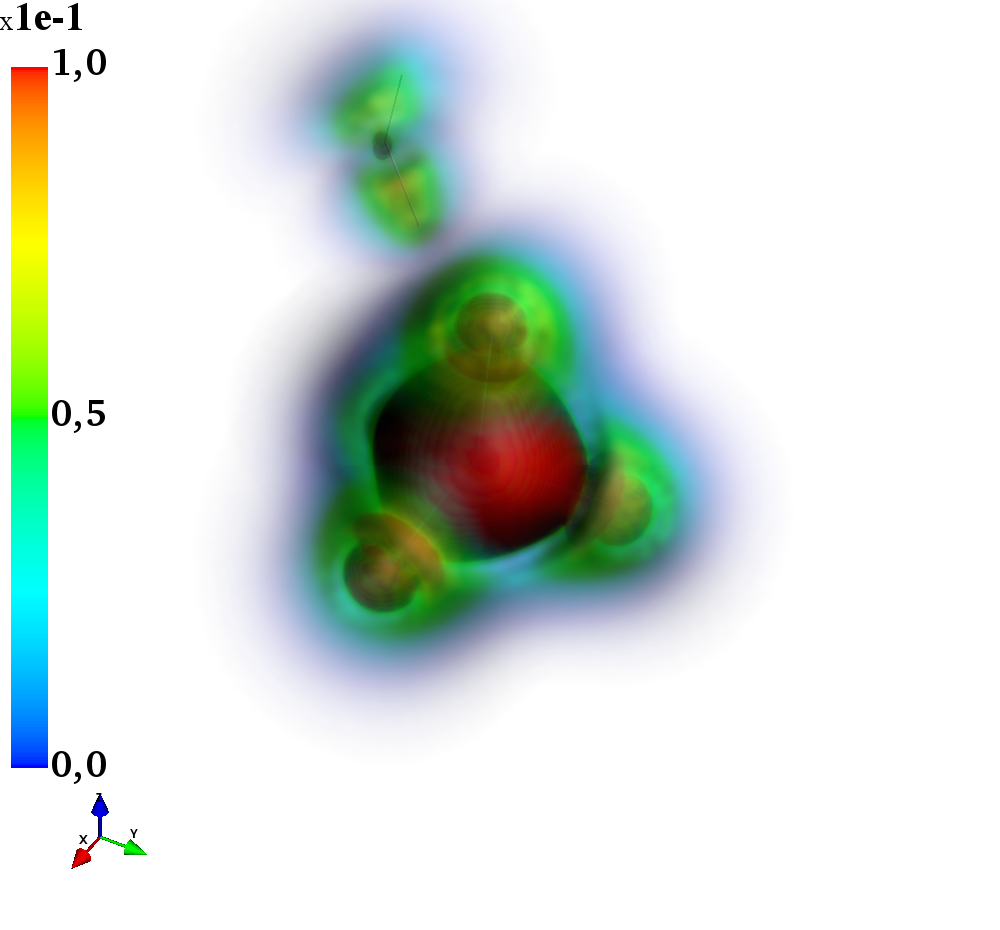}
		\label{graph:6a_p1}
	\end{minipage}\hfill
	\begin{minipage}[t]{0.33\linewidth}
	\includegraphics[width=\linewidth]{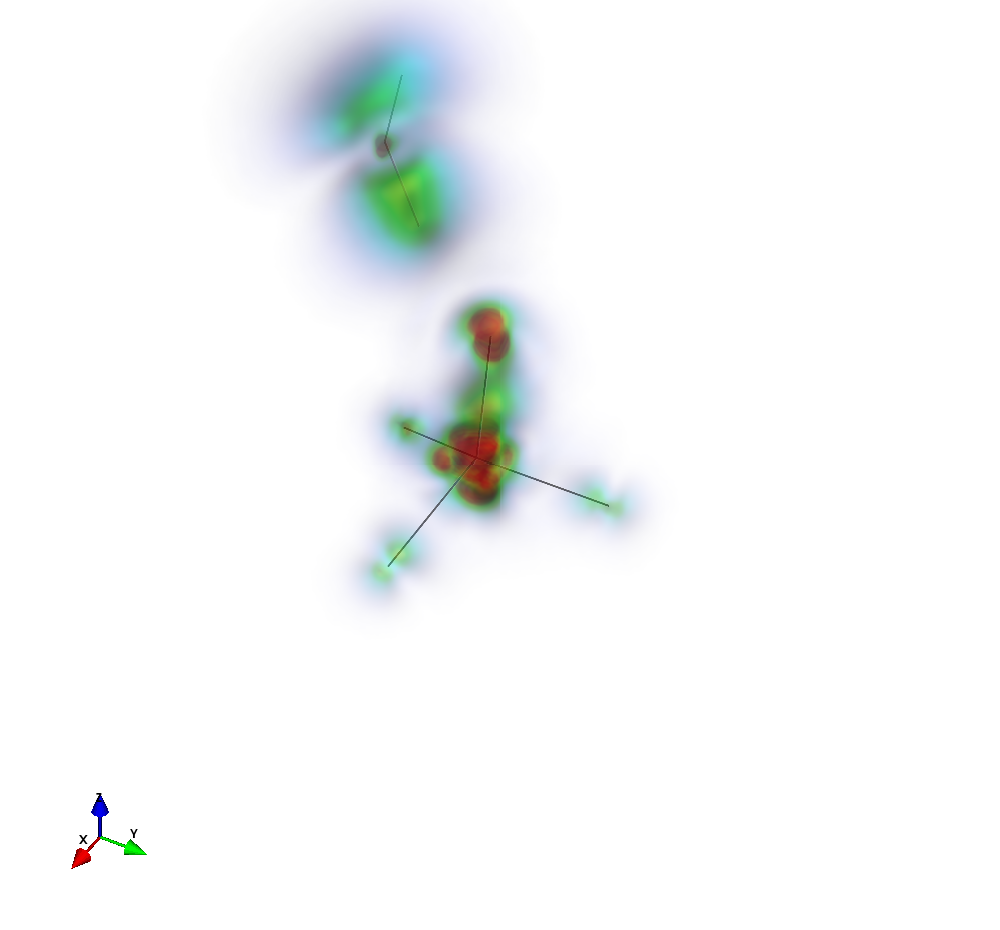}
	\label{graph:6b_p1}
\end{minipage}\hfill
	\begin{minipage}[t]{0.33\linewidth}
	\includegraphics[width=\linewidth]{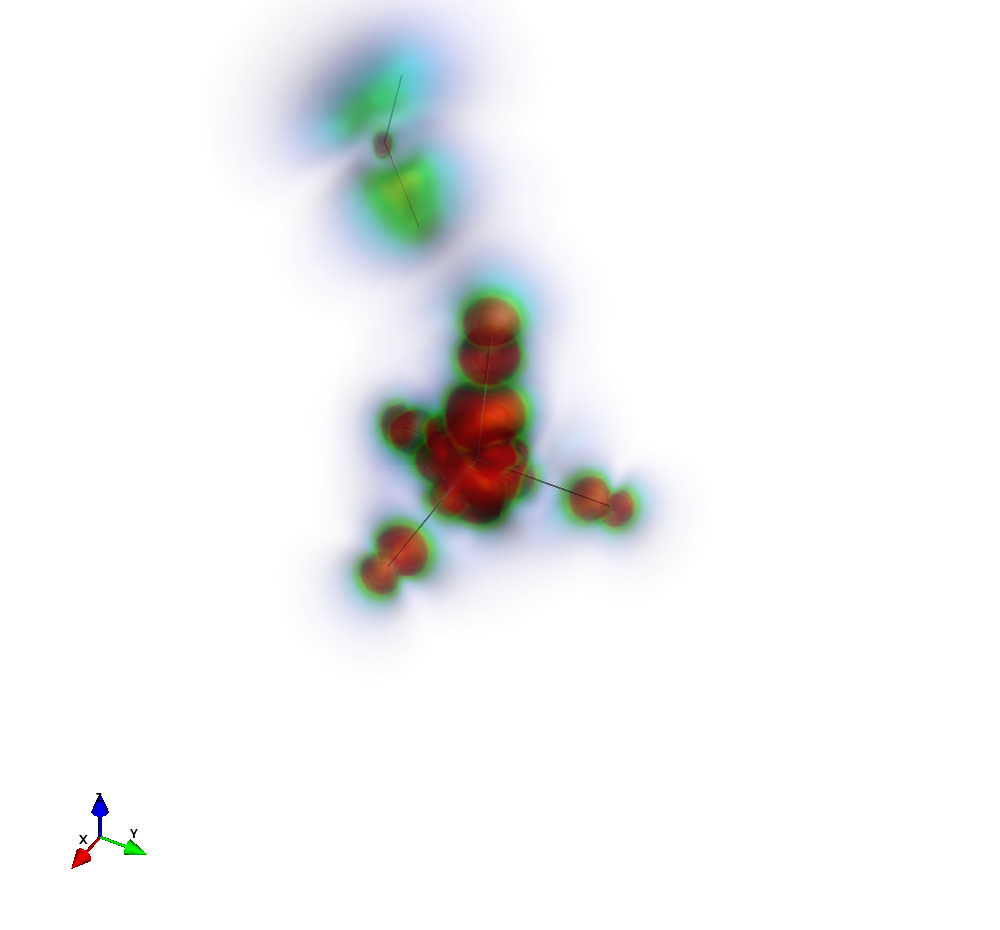}
	\label{graph:6c_p1}
\end{minipage}\hfill
\vspace{-1\baselineskip}
p = 3\\%
	\begin{minipage}[t]{0.33\linewidth}
	\includegraphics[width=\linewidth]{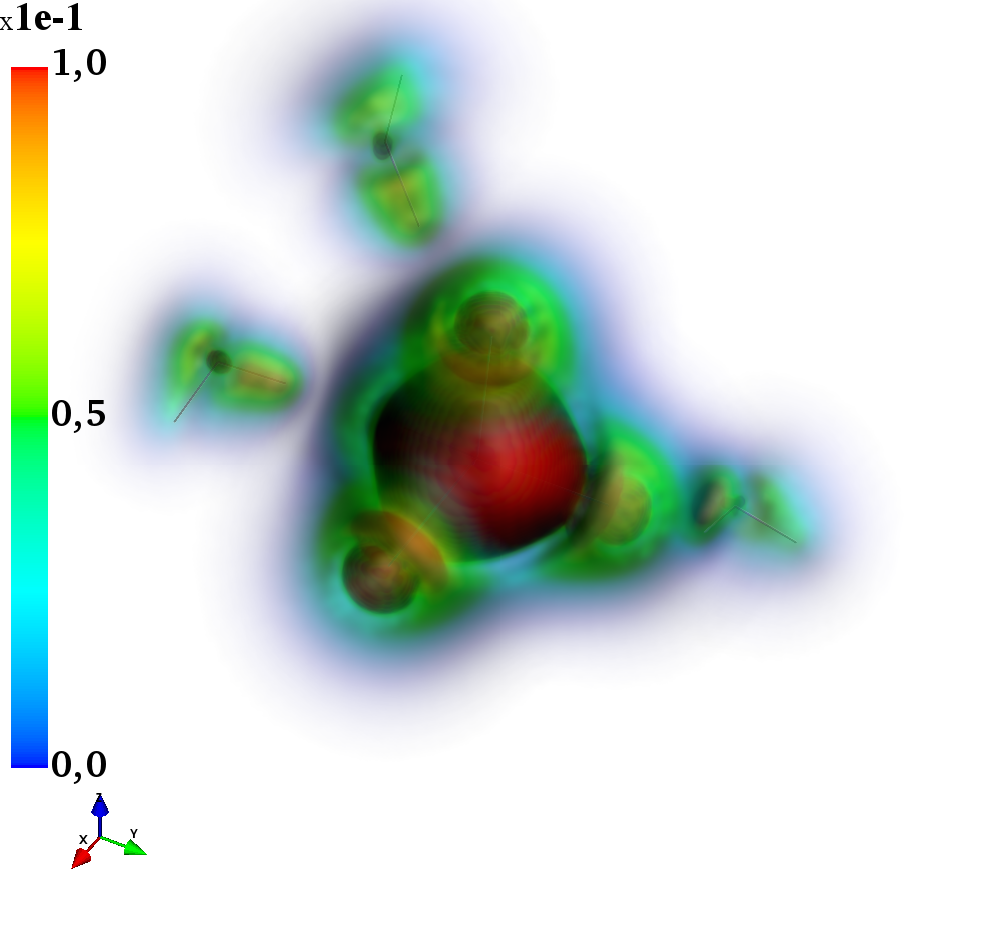}
	\label{graph:6a_p3}
\end{minipage}\hfill
\begin{minipage}[t]{0.33\linewidth}
	\includegraphics[width=\linewidth]{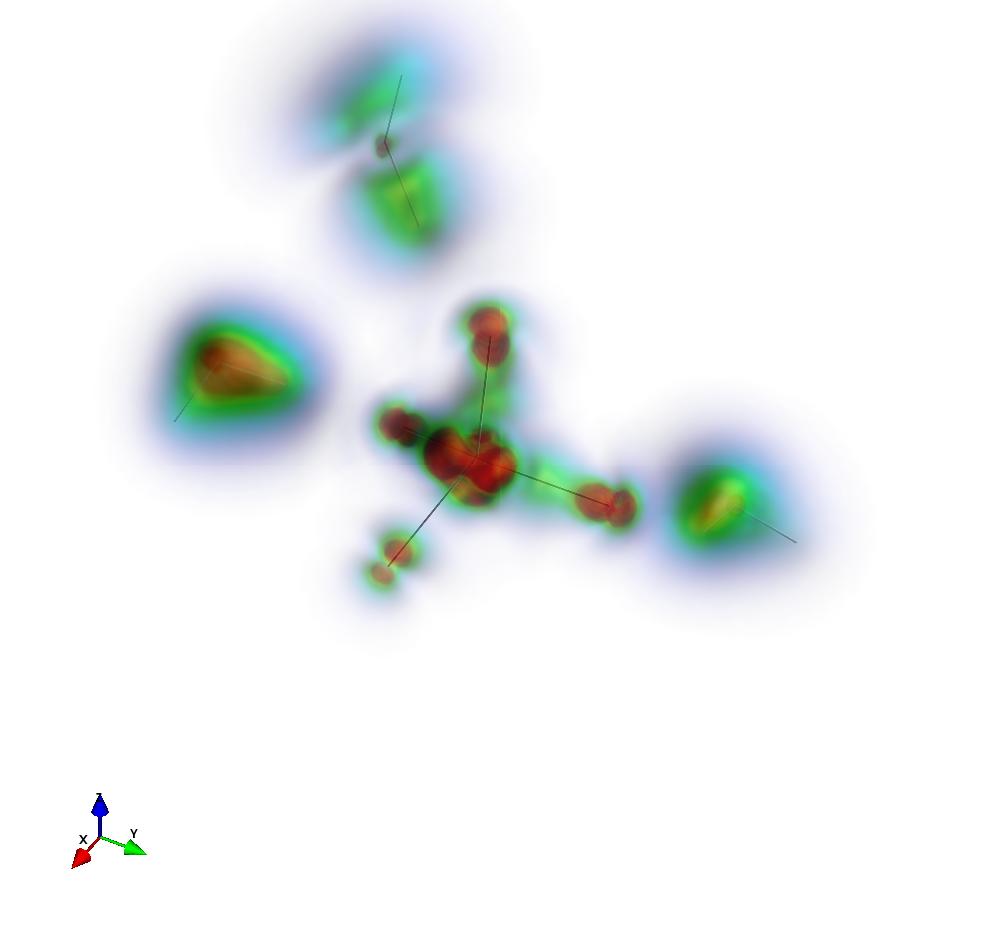}
	\label{graph:6b_p3}
\end{minipage}\hfill
\begin{minipage}[t]{0.33\linewidth}
	\includegraphics[width=\linewidth]{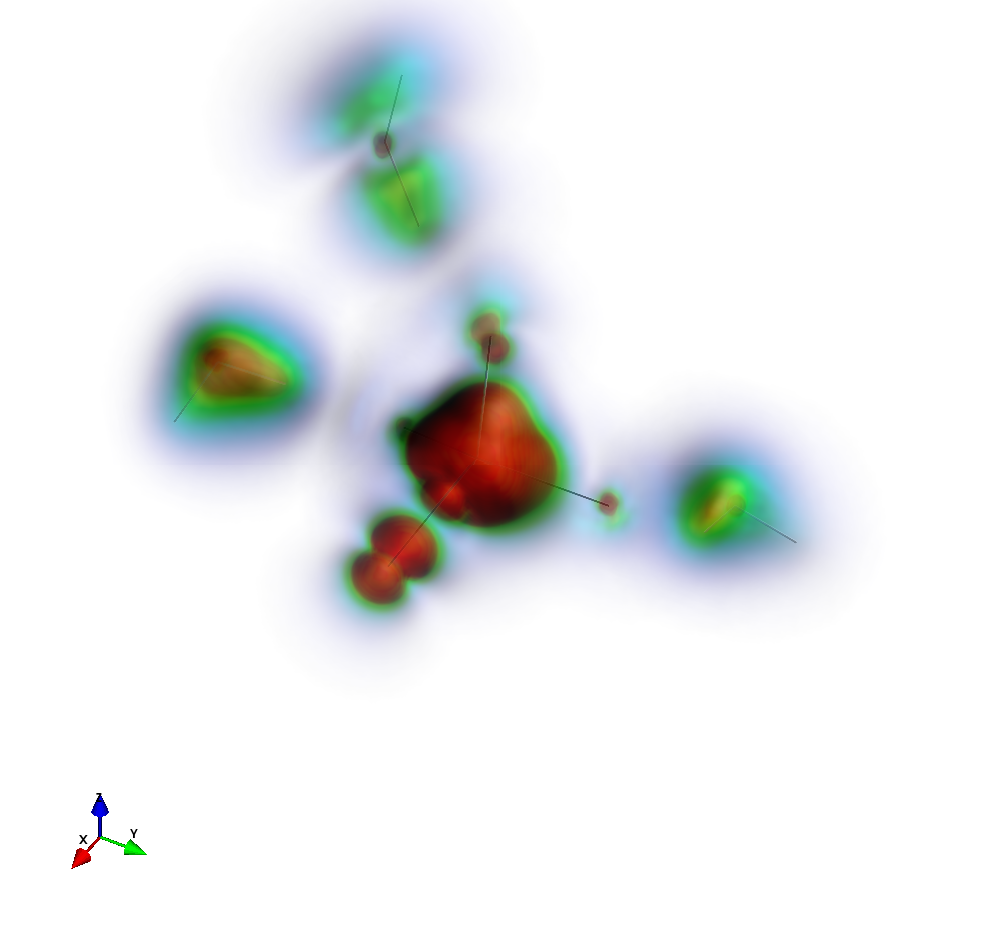}
	\label{graph:6c_p3}
\end{minipage}\hfill
\vspace{-1\baselineskip}
p = 5\\%
\begin{minipage}[t]{0.33\linewidth}
	\includegraphics[width=\linewidth]{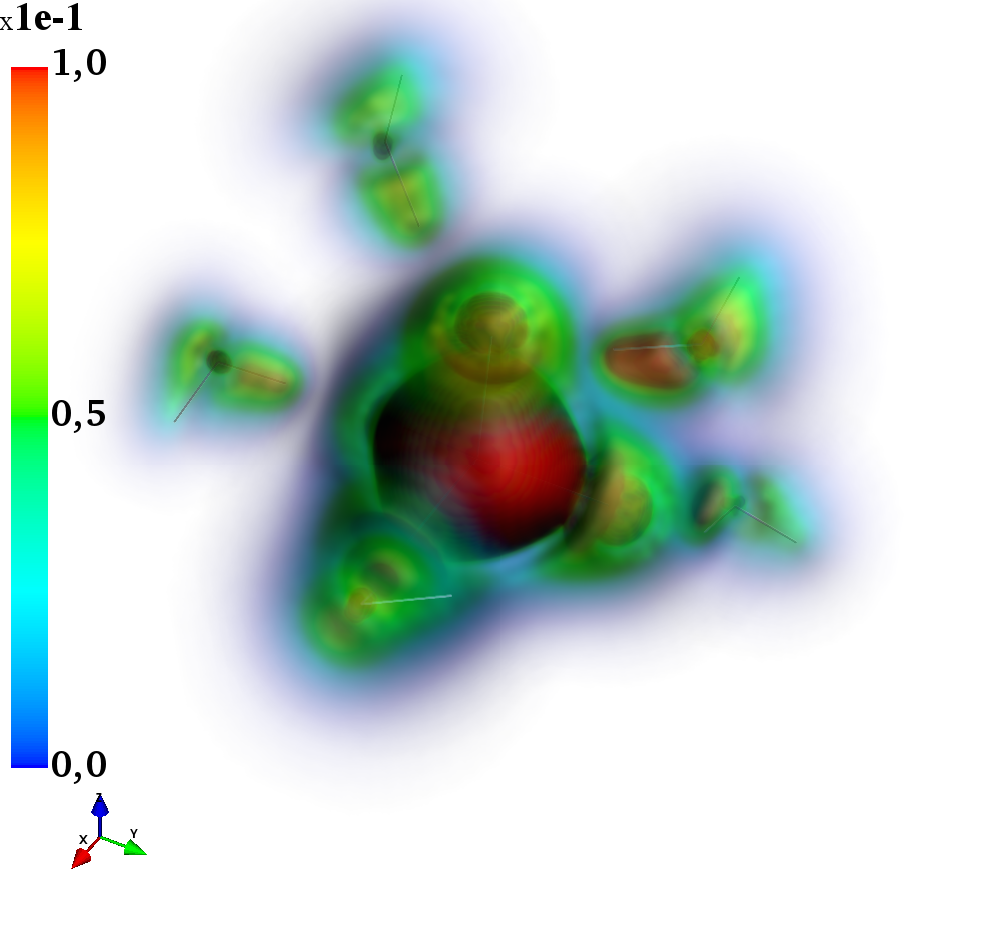}
	\label{graph:6a_p5}
\end{minipage}\hfill
\begin{minipage}[t]{0.33\linewidth}
	\includegraphics[width=\linewidth]{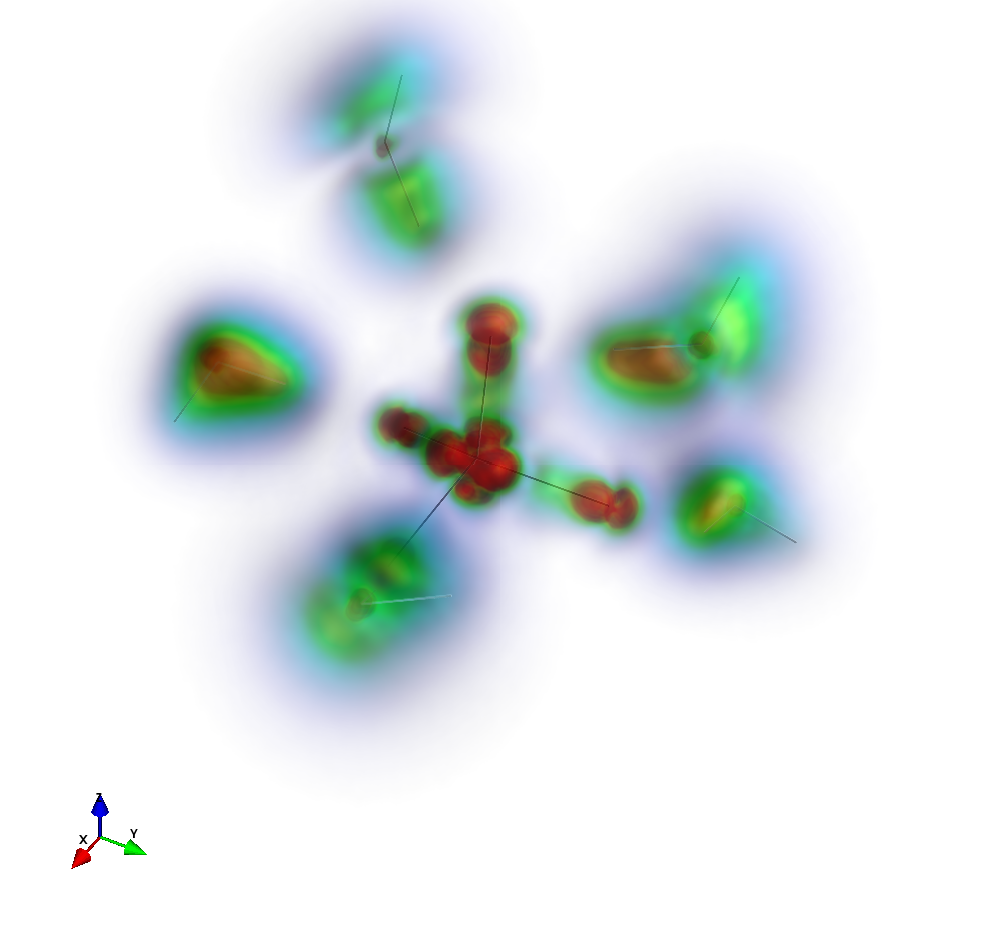}
	\label{graph:6b_p5}
\end{minipage}\hfill
\begin{minipage}[t]{0.33\linewidth}
	\includegraphics[width=\linewidth]{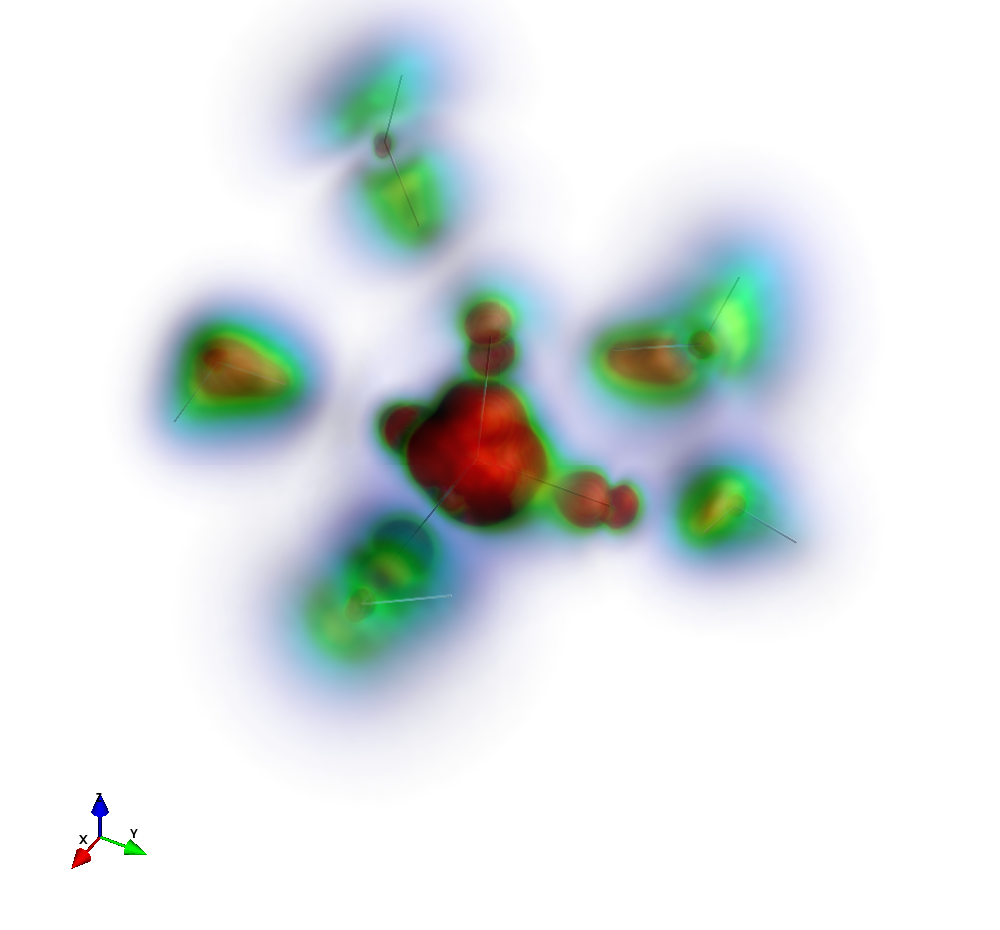}
	\label{graph:6c_p5}
\end{minipage}\hfill%
	\caption{Volumetric plots of shielding densities and differences in 
	shielding densities between 0.0 and 0.1 ppm, computed with the DC 
	Hamiltonian, for \ce{[MoO4(H2O)_p]^{2-}} with increasing 
	number of $p$ water molecules ($p$ =1, 3, 5). Left column: 
	Supermolecule ($[A+pB|0|0]$); Middle column: difference between 
	supermolecular and electronic embedding ($[A|6B_g|0]$) calculations; 
	Right column: difference between supermolecular and mechanical 
	embedding ($[A|0|0]+[0|6B_g|0]$) calculations. For embedded models, 
	water molecule contributions to \ce{^{95}Mo} shielding densities are 
	calculated with the NICS method.}
	\label{graph:volumetric_diff_all_dirac_p_1_3_5}
\end{figure}

\end{document}